\documentclass[twocolumn]{aastex631}
\usepackage{xcolor}
\usepackage{threeparttable}
\usepackage{amsmath}

\usepackage{booktabs}
\shorttitle{High-Energy Excess of Swift J1727.8-1613}
\shortauthors{Xiao Fan}
\graphicspath{{./}}

\begin{document}

\title{Evolution of the High-Energy Excess in Swift J1727.8-1613 during Its 2023 Flare State}

\correspondingauthor{Bei You}
\email{youbei@whu.edu.cn}

\author[0000-0001-7350-8380]{Xiao Fan}
\author[0000-0002-8231-063X]{Bei You}
\affiliation{Department of Astronomy, School of Physics and Technology, Wuhan University, Wuhan 430072, People's Republic of China}
\author[0009-0007-1966-181X]{Yi Long}
\affiliation{Department of Astronomy, Nanjing University, 163 Xianlin Avenue, Nanjing 210023, Peoples Republic of China}
\author[0009-0007-7292-8392]{Han He}
\affiliation{Department of Astronomy, School of Physics and Technology, Wuhan University, Wuhan 430072, People's Republic of China}
\author[0000-0002-1908-0536]{Liang Chen}
\affiliation{Shanghai Astronomical Observatory, Chinese Academy of Sciences (CAS), Shanghai 200030, People’s Republic of China}

\begin{abstract}

We investigate the evolution of the high-energy excess in Swift J1727.8-1613 during its 2023 flare state using joint Insight-HXMT and INTEGRAL/ISGRI observations covering 2--500 keV. Fits to the broadband spectra with a model including thermal Comptonization and disk reflection leave systematic positive residuals above $\sim150\,{\rm keV}$. Adding a power-law component improves the fits and yields photon indices of $\Gamma_{\rm PL}=2.08$--2.61, demonstrating that the high-energy excess beyond the thermal Comptonization continuum is present throughout the flare state. We also show that all spectra can be well described by the hybrid Comptonization model \texttt{eqpair}, without requiring any additional high-energy component. The fitting results indicate that the energy supply to the electrons transitions from predominantly thermal heating to non-thermal acceleration, and that the electron injection spectrum steepens abruptly near the onset of the flare state. During this evolution, the radio spectral index changes from approximately zero to negative values, accompanied by the ejection of three transient jet knots. Based on this temporal association, we discuss a possible connection between the increase in non-thermal power indicated by our spectral analysis and the ejection of the transient jet.

\end{abstract}

\keywords{Stellar mass black holes; Accretion; X-ray binaries}

\section{Introduction}
Low-mass X-ray binaries (LMXBs) are binary systems in which a compact object (e.g., a black hole in this work) accretes material from a low-mass companion star ($\lesssim 1 \, M_{\odot}$) via Roche lobe overflow, producing broadband emission spanning from radio to X-ray \citep[see][for review]{Belloni2016}. The X-ray spectra of LMXBs generally consist of thermal emission from an optically thick accretion disk \citep{SSD1973,Mitsuda1984} and a harder Comptonization component arising from a hot accretion flow or a Comptonizing corona \citep{Sunyaev1980,Zdziarski2004}. Radio emission of LMXBs is commonly interpreted as synchrotron radiation from a compact relativistic jet \citep{Hjellming1988,2006csxs.book..381Fender}. During an outburst, black hole LMXBs typically undergo a state transition \citep[see][for a review]{Remillard2006ARAA}. They evolve from the hard state, dominated by Comptonized emission, through the hard intermediate state (HIMS) to the soft state, dominated by the emission of the accretion disk, and eventually return to the hard state, tracing a ``q-shaped" track in the hardness--intensity diagram \citep[e.g.,][]{Homan2001,Homan2005state}. The steady compact jet is present in the hard state and is suppressed as the source evolves into the soft state; it reappears when the source re-enters the hard state \citep{corbel2013,bright2020,you2023,you2024}. This is thought to be associated with the magnetic field transport via the accretion
\citep{du2026}. Bright radio flares, sometimes resolved into discrete relativistic ejecta, are often observed during the hard-to-soft transition and generally exhibit optically thin synchrotron emission, which is characterized by a negative radio spectral index \citep{Fender2004,Fender2009,carotenuto2021,cao2025}.

At high energies, the thermal Comptonization spectrum is characterized by a high-energy cutoff related to the electron temperature. However, an additional emission component extending beyond this cutoff has been detected above $\sim200\,{\rm keV}$ in several black hole X-ray binaries, which commonly be referred to as a ``high-energy excess" or ``high-energy tail" \citep[e.g.,][]{McConnell2002ApJ,DelSanto2008,Bouchet2009,Roques2015,Roques2019,Bassi2020,cangemi2021}. Recently, \citet{yang2026} studied the covariance spectrum of MAXI J1820+070 in 2-150 keV during its hard state. A clear decrease in coherence above about 30 keV with respect to the reference band of 2-10 keV is observed, implying that the hard X-ray variability at high energies is not fully correlated with the soft X-ray variability. It suggests that the observed X-ray variability cannot be explained by a single, uniform Comptonization component. Instead, at least two variability components are required: one component is coherent with the 2–10 keV band and dominates below tens of keV, while another component becomes increasingly important above 30 keV and varies incoherently with the soft band.

Neglecting high-energy excess may lead to an overestimate of the electron temperature, and the variations in the high-energy flux may also be misinterpreted as changes in the electron temperature or optical depth of the thermal Comptonizing plasma \citep{Droulans2010}. In fact, the excess varies while the spectral shape of the thermal Comptonization continuum remains nearly unchanged, suggesting that the two components may have different origins \citep{Droulans2010}. However, the physical origin of the high-energy excess remains unclear \citep[e.g.,][]{Jourdain2017,Bassi2020}. Proposed explanations include an additional Comptonizing region or a temperature gradient within the plasma \citep[e.g.,][]{Bouchet2009}, Comptonization by a hybrid thermal/non-thermal electron population \citep[e.g.,][]{Coppi1999,McConnell2002ApJ,DelSanto2008,Bassi2020,2021ApJ_Zdziarski_hybrid}, and synchrotron or synchrotron self-Compton emission from the hot flow and/or the jet \citep[e.g.,][]{Laurent2011,veledina2011,Jourdain2012jet,Zdziarski2017,yang2026}.

Swift J1727.8-1613 is a recently discovered LMXB, first detected by Swift/BAT on 2023 August 24 \citep[MJD 60180;][]{2023GCNPage} and subsequently identified as an LMXB through multiwavelength follow-up observations \citep{2023ATelNakajima, 2023ATelCastro-Tirado, 2023ATelMiller-Jones, 2023ATelOConnor} and optical dynamical studies \citep{2025A&AMataSanchez}. With a peak flux of $\sim$8 Crab in the 15--50 keV band \citep{2023ATelPalmer}, Swift J1727.8-1613 is one of the brightest X-ray binary ever recorded, making it an exceptional object for studying accretion and jet physics across radio, optical, and X-ray wavelengths \citep[e.g.,][]{veledina2023,chatterjee2024,ingram2024,nandi2024,Shui2024,Yang2024,zhu2024,jin2025,Liao2025,ma2025,rawat2025,Peng2025,vincentelli2025,Xusaien2025ApJ,Zdziarski2025,li2026,ma2026,ma2026mnras,nitindala2026,yu2026}. A notable feature of the 2023 outburst of Swift J1727.8-1613 is its unusually prolonged HIMS lasting $\sim$43 days \citep{2025arXiv_He}, during which the source entered a so-called ``flare state'' \citep{2024MNRASYu} characterized by successive soft X-ray flares while the hard X-ray emission declined monotonically. \citet{2025arXiv_He} attributed these flares to propagating fluctuations in the accretion disk, which may be driven by thermal-viscous instabilities, and demonstrated this interpretation quantitatively using the propagating fluctuation model. Meanwhile, radio observations revealed multiple bright flares and transient jet activity during the flare state, including the ejection of three discrete knots \citep{Hughes2025,Wood2025}. \citet{Xusaien2025ApJ} presented a timing and spectral analysis of Swift J1727.8-1613, in which the correlation of QPO frequency with the spectral components was investigated. A novel two-branch correlation between QPO frequency and observed disk emission was discovered: below \~ 3 Hz the frequency is negatively correlated with disk luminosity, while above \~ 3 Hz the relation transitions to positive. The QPO frequency versus Compton flux shows the opposite trend, indicating a systematic interaction between the disk and the corona.

The emission above $200 \,{\rm keV}$ from Swift J1727.8-1613 was first reported by \citet{Cangemi2023ATel}, and an additional high-energy excess beyond the thermal Comptonization continuum has since been confirmed by several studies \citep[e.g.,][]{Mereminskiy2024,2024ApJPengJQ,2024A&A_Bouchet,Caojia2025,Chand2026,Liu2026}. Its physical origin remains debated, with the main interpretations being Comptonization by a hybrid thermal/non-thermal electron population and emission from the jet. Based on the hard-state and early-HIMS spectra, both \citet{Liu2026} and \citet{Chand2026} found that purely thermal Comptonization cannot reproduce the high-energy emission, whereas models including a hybrid thermal/non-thermal electron population provide acceptable fits. On the other hand, \citet{2024A&A_Bouchet} detected highly polarized emission above 210 keV during the HIMS. Although the polarization angle was not aligned with the projected direction of the compact jet, the high degree of polarization nevertheless suggests that the jet may also contribute to the high-energy excess. A jet origin was also suggested by \citet{Caojia2025}, who attributed the steep power-law component to synchrotron self-Compton emission based on its weak reflection and lack of correlation with the disk emission. These two interpretations are not necessarily mutually exclusive. \citet{Mereminskiy2024} and \citet{2024ApJPengJQ} noted that the available spectra could not distinguish conclusively between hybrid Comptonization and jet emission, while \citet{Peng2025} and \citet{Caojia2025} proposed geometries in which both the hot inner flow and jet contribute to the observed spectrum.

In this work, we combine Insight-HXMT and INTEGRAL/ISGRI observations to extend the spectral coverage to $2$--$500\,{\rm keV}$, allowing us to better constrain the spectral properties and evolution of the high-energy excess during the flare state. The observations and data analysis are described in Section~\ref{sec:2}. In Section~\ref{sec:3}, we establish the presence of the high-energy excess and investigate its spectral evolution using a hybrid Comptonization model. The results are discussed and summarized in Section~\ref{sec:4}.

\section{Observations and Data Analysis}\label{sec:2}

\begin{figure*}[htbp]
    \centering
    \includegraphics[scale=0.55]{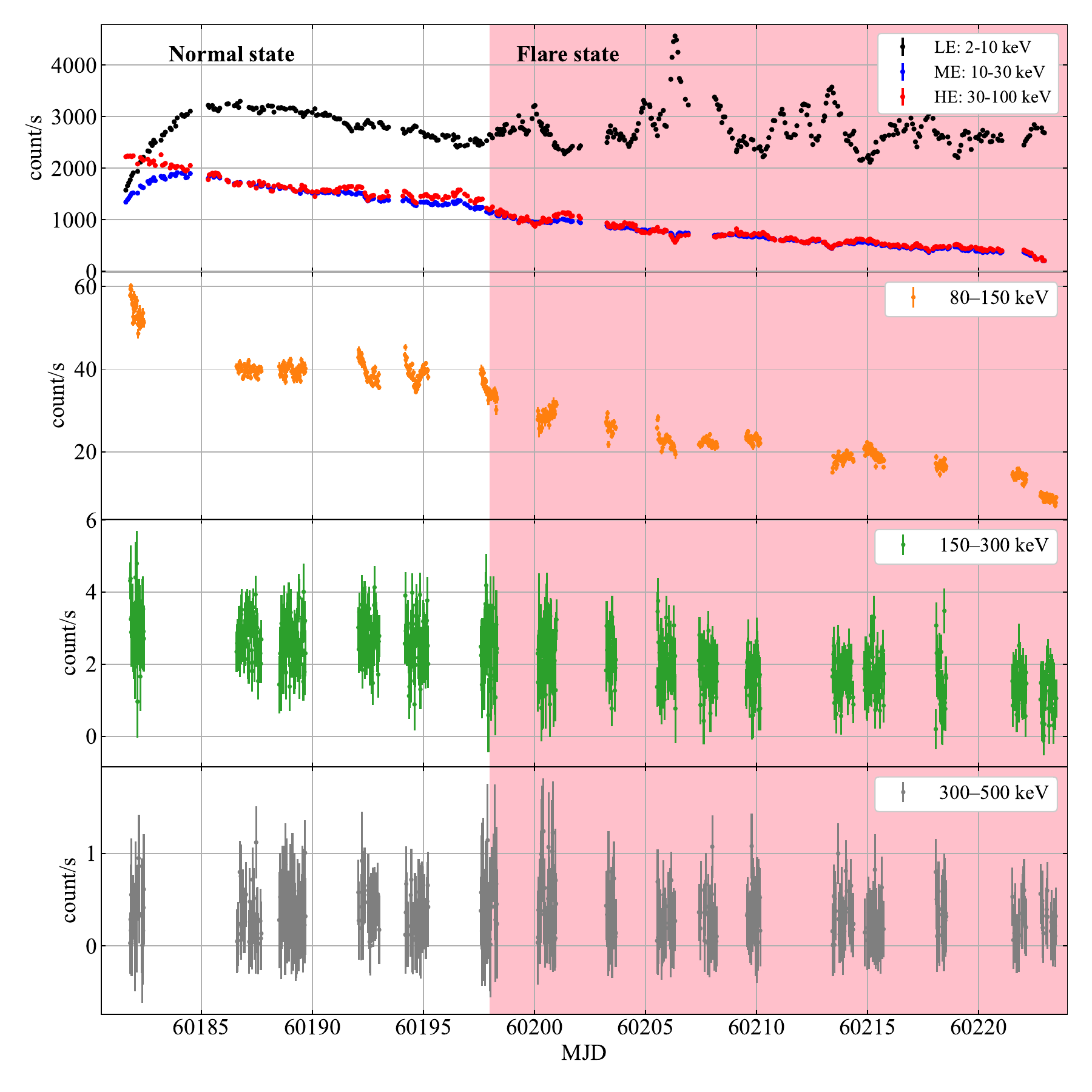}
    \caption{Light curves of Swift J1727.8-1613 during the 2023 outburst. \textit{Top panel}: Light curves obtained with the three Insight-HXMT instruments: LE (2--10 keV, black), ME (10--30 keV, blue), and HE (30--100 keV, red). \textit{Lower three panels}: Light curves obtained with INTEGRAL/ISGRI in three energy bands: 80--150 keV (orange), 150--300 keV (green), and 300--500 keV (gray). The pink shaded region indicates the flare state.}
    \label{fig:lightcurve}
\end{figure*}

\subsection{Insight-HXMT}
Insight-HXMT is China's first X-ray astronomical satellite launched on 2017 June 15, carrying three main instruments with different effective areas \citep{HXMT-Zhang2020}: the Low Energy telescope (LE, $384 \, {\rm cm^2}$ at 1--12 keV), the Medium Energy telescope (ME, $952\, {\rm cm^2}$ at 8--35 keV), and the High Energy telescope (HE, $5100 \, {\rm cm^2}$ at 20--250 keV). The three instruments provide broad energy coverage for X-ray spectral studies of X-ray binaries \citep[e.g.,][]{daixiaohang2023MNRAS,2024ApJPengJQ,Caojia2025,Fanxiao2026ApJ}. 

Swift J1727.8-1613 was monitored by Insight-HXMT from 2023 August 25 (MJD 60181) to 2023 October 6 (MJD 60223) with high cadence. Data from all three instruments were processed using the Insight-HXMT Data Analysis Software (HXMTDAS, v2.06) with the default configuration. The adopted energy bands in this work are 2--10 keV (LE), 10--30 keV (ME, excluding 21--24 keV due to silver fluorescence lines) and 30--100 keV (HE). A systematic uncertainty of 1.5\% was added to the spectra from each instrument to account for residual calibration uncertainties \citep[e.g.,][]{2021NatCo_You}. Joint LE, ME, and HE spectra were extracted based on good time intervals (GTIs). Following \citet{2025arXiv_He}, the ME GTIs were used as the time reference because they generally encompass those of LE and HE. GTIs with short exposure times were merged with adjacent ones to improve the signal-to-noise ratio.

The light curves obtained with the three instruments are shown in Figure~\ref{fig:lightcurve}. Following \citet{2024MNRASYu}, we divide the outburst into a normal state before MJD~60198 and a flare state thereafter. During the flare state, the soft X-ray flux shows a series of prominent flares while the hard X-ray flux generally decreases.

\subsection{INTEGRAL/ISGRI}

The INTEGRAL Soft Gamma-Ray Imager \citep[ISGRI;][]{2003A&ALebrun} is the CdTe detector plane of the coded-mask telescope IBIS aboard INTEGRAL \citep{Winkler2003}, covering the 15--1000 keV energy range. During the Insight-HXMT monitoring period, INTEGRAL observed Swift J1727.8-1613 from satellite revolutions 2678--2694. The observations consist of a series of individual pointings, referred to as Science Windows (ScWs).

INTEGRAL data were reduced using the INTEGRAL Off-line Scientific Analysis software \citep[OSA v11.2;][]{2003A&ACourvoisier} following standard procedures \footnote{\url{https://www.isdc.unige.ch/integral/download/osa/doc/11.2/osa_um_ibis/man_html.html}}. For each ScW, sky images were reconstructed, and source fluxes were extracted by deconvolving the coded-mask shadowgrams. The long-term degradation of the CdTe detectors results in increasing calibration uncertainties at low energies \citep{2024A&A_Bouchet,Mereminskiy2024}. To evaluate the effect of this degradation, we compared ISGRI spectra of the Crab obtained early in the mission (revolution 39) and close to our observations (revolution 2686). The recent Crab spectrum shows clear deviations at low energies, while the two spectra agree above 40 keV. We therefore restricted the ISGRI analysis to the 40--500 keV band and reconstructed sky images in the 40--80, 80--150, 150--300, and 300--500 keV bands.

Light curves were extracted with a time bin of $2 \, {\rm ks}$ and the results are shown in Figure~\ref{fig:lightcurve}. The 40--80 keV light curve is not displayed because its energy range overlaps with that of Insight-HXMT/HE. Sources with a detection significance above $7 \sigma$ in the reconstructed sky images were retained for subsequent spectral extraction. Spectra were extracted from 40 keV to 500 keV, and a systematic uncertainty of 3\% was added to each spectral channel, following the recommendation of the OSA User Manual.

\subsection{Spectral Analysis} \label{sec:spectral_analysis}

\begin{figure}[htbp]
    \centering
    \includegraphics[scale=0.34]{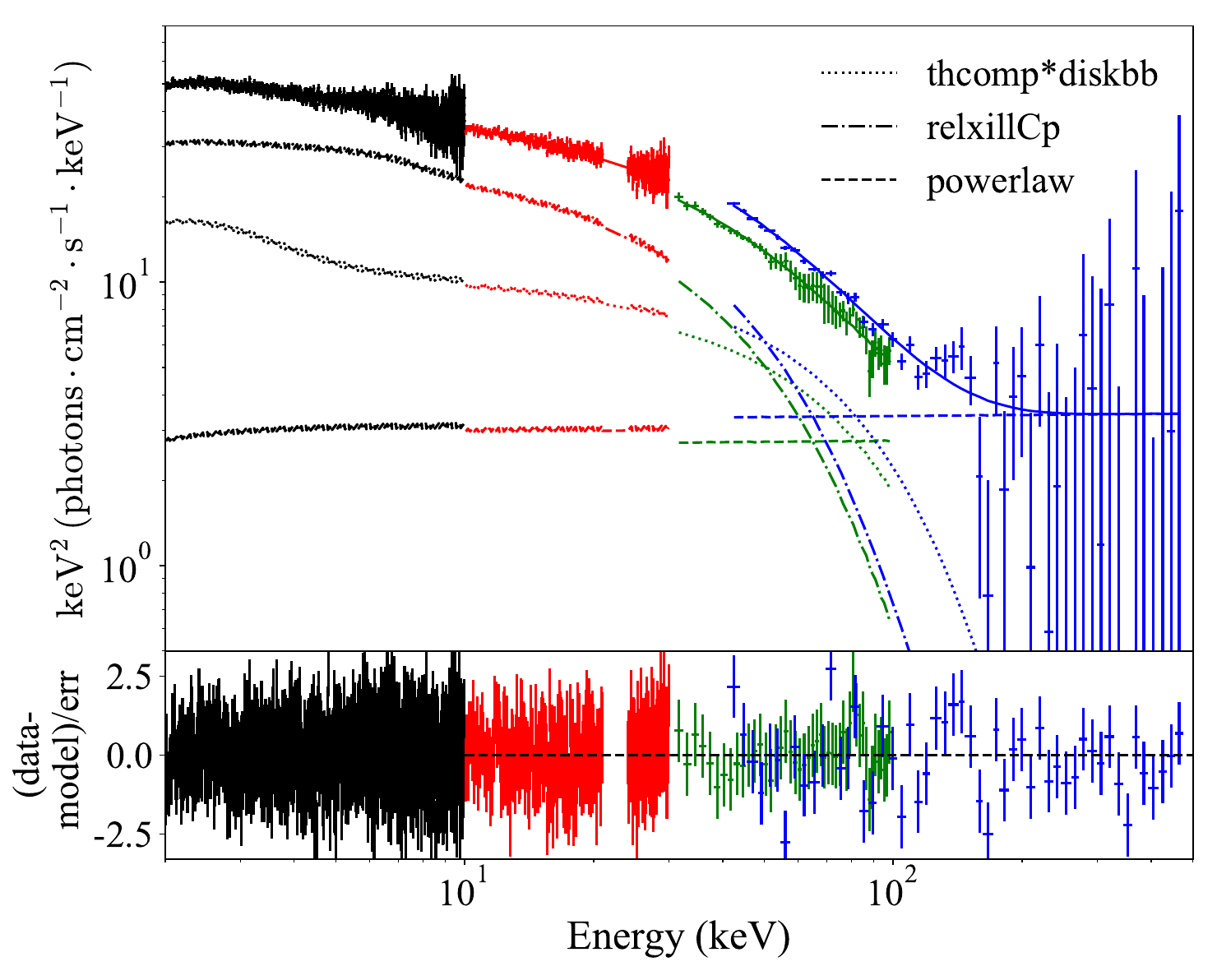}
    \caption{A representative simultaneoust Insight-HXMT and INTEGRAL spectrum on MJD~60200 used for cross-calibration, fitted with \texttt{constant*tbabs*(thcomp*diskbb+relxillCp+powerlaw)}. The LE, ME, HE, and ISGRI data are shown in black, red, green, and blue, respectively. The lower panel shows the residuals in units of the statistical uncertainty.}
    \label{fig:cross-calibration}
\end{figure}

To investigate the high-energy spectral behavior of Swift J1727.8-1613 during the flare state, we performed a broadband spectral analysis by combining Insight-HXMT and INTEGRAL/ISGRI data over the 2--500 keV band. Because the Insight-HXMT/LE data before MJD$\sim$60200 were affected by pile-up and subject to significant calibration uncertainties \citep[also see][]{2024ApJPengJQ,Peng2025}, we analyzed only the broadband spectra obtained after this time, beginning with INTEGRAL revolution 2685. It has been suggested that systematic flux differences between different instruments are common and can reach tens of percent \citep[e.g.,][]{2017AJ_Madsen}. A multiplicative constant is therefore included in joint spectral fitting to account for this effect \citep[e.g.,][]{2021NatCo_You, Caojia2025, 2024ApJPengJQ, 2025arXiv_He}. In this work, a constant was included for each instrument relative to LE, which was fixed at unity. The overlapping 40--100 keV coverage of Insight-HXMT/HE and INTEGRAL/ISGRI was used to examine the cross-calibration between the two instruments.

\begin{figure*}[htbp]
    \centering
    \includegraphics[scale=0.38]{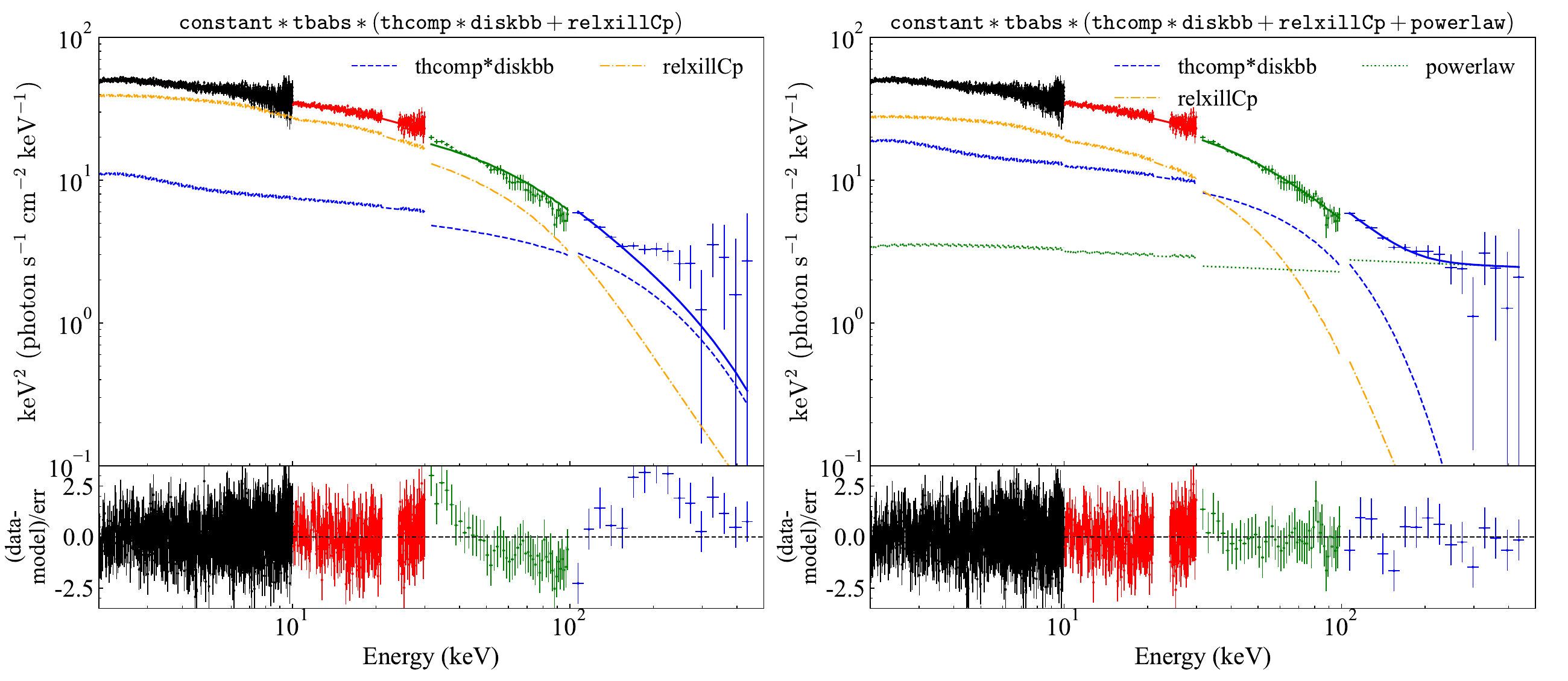}
    \caption{Representative fitting results of joint Insight-HXMT and INTEGRAL spectrum at INTEGRAL revolution 2685. The left and right panels show the residuals obtained with the baseline model and after adding a power-law component, respectively.}
    \label{fig:excess}
\end{figure*}

We selected a representative joint spectrum on MJD~60200, consisting of an Insight-HXMT observation (exposure time of $\sim400 \,$s) whose GTI falls entirely within a contemporaneous ISGRI observation (exposure time of $\sim1840 \,$s). We first fitted the joint spectrum with the model \texttt{constant*tbabs*}\allowbreak\texttt{(thcomp*diskbb+}\allowbreak\texttt{relxillCp}\allowbreak\texttt{+powerlaw)}, following \citet{2025arXiv_He}. Throughout the spectral analysis, the hydrogen column density was fixed at $N_{\rm H}=0.226\times10^{22}\,{\rm cm}^{-2}$. In the fits including \texttt{relxillCp}, the black hole spin and disk inclination were fixed at $0.98$ and $40^\circ$, inferred from joint broadband spectral fitting of NICER, NuSTAR, and Insight-HXMT data with relativistic reflection models by \citet{2024ApJPengJQ}. The inner disk radius was linked to the \texttt{diskbb} normalization rather than being fitted independently \citep[see Section 3.2 of][]{2025arXiv_He}. The photon index and electron temperature of \texttt{relxillCp} were tied to those of \texttt{thcomp}, while the reflection fraction was fixed at $-1$ so that \texttt{relxillCp} included only the reflected emission. The iron abundance was fixed at the solar value.

We obtained best-fitting cross-normalization constants relative to LE of $0.963 \pm 0.007$ (ME), $0.826 \pm 0.021$ (HE), and $1.004 \pm 0.025$ (ISGRI) respectively. As shown in Figure~\ref{fig:cross-calibration}, the Insight-HXMT/HE and INTEGRAL/ISGRI data show good agreement in the overlapping 40--100 keV band. This agreement supports the joint use of the two satellites for broadband spectral analysis. In the subsequent analysis, only the ISGRI data above 100 keV were used to avoid overlap with Insight-HXMT/HE. 

\begin{table*}
\centering
\caption{Summary of INTEGRAL/ISGRI and Insight-HXMT observations used for broadband spectral analysis. The HXMT exposure time refers to the GTI duration.}
\label{tab:obs}
\begin{tabular}{cccccc}
\hline\hline
INTEGRAL Rev. & INTEGRAL MJD & ISGRI Exp.& HXMT Obs. ID &HXMT MJD &HXMT/ME Exp.\\
\# & & (ks) & & & (s) \\
\hline
2685 & 60200.13--60201.00 & 47.8 & P061433801404 & 60200.52 & 360.8 \\
2686 & 60203.22--60203.67 & 26.6 & P061433801704 & 60203.46 & 535.2 \\
2687 & 60205.51--60206.35 & 48.9 & P061433801907 & 60205.94 & 344.2 \\
2689 & 60209.52--60210.17 & 37.3 & P061433802306 & 60209.83 & 641.2 \\
2690 & 60213.41--60214.37 & 55.0 & P061433802705 & 60213.66 & 965.8 \\
2691 & 60214.84--60215.75 & 52.5 & P061433802902 & 60215.24 & 901.9 \\
2692 & 60218.06--60218.55 & 28.2 & P061433803203 & 60218.41 & 1106.0 \\
2693 & 60221.51--60222.16 & 33.9 & P061433803501 & 60222.04 & 867.0 \\
2694 & 60222.90--60223.46 & 40.9 & P061433803507 & 60222.89 & 938.1 \\
\hline
\end{tabular}
\end{table*}

As illustrated in Figure~\ref{fig:cross-calibration}, individual ScW spectra do not have sufficient signal-to-noise ratios above $\sim 100\, $keV. Since the hard X-ray variability of Swift J1727.8-1613 within each INTEGRAL revolution is small (see Figure~\ref{fig:lightcurve}), we combined all ScW spectra within each revolution using the \texttt{spe\_pick} tool to obtain average spectra with improved signal-to-noise ratio. The resulting revolution-averaged spectra have effective exposure times of 30--$60\, {\rm ks}$. In contrast, the soft X-ray flux of Swift J1727.8-1613 varies significantly on timescales of hundreds of seconds \citep{2025arXiv_He}; we therefore used individual Insight-HXMT GTI spectra rather than averaging them. Each revolution-averaged ISGRI spectrum was then paired with a contemporaneous Insight-HXMT GTI spectrum to construct a 2--$500 \, {\rm keV}$ broadband spectrum. In total, nine joint Insight-HXMT and INTEGRAL broadband spectra were constructed, and the corresponding observation information is summarized in Table~\ref{tab:obs}. The resulting broadband spectra are analyzed below using both phenomenological models and the hybrid Comptonization model \texttt{eqpair} to investigate the high-energy excess.

\section{Results} \label{sec:3}
\subsection{The High-energy excess of Swift J1727.8-1613}

\begin{table*}
\centering
\caption{Results of the phenomenological fits to spectra of Swift JJ1727.8-1613 during the flare state. 
The baseline model is \texttt{constant*tbabs*(thcomp*diskbb+relxillCp)},
while the power-law model includes an additional \texttt{powerlaw} component. 
The improvement in the fit, $\Delta\chi^2$, corresponds to two additional free parameters.}
\label{tab:powerlaw}
\begin{tabular}{lcccc}
\hline\hline
Revolution 
& $\Gamma_{\rm PL}$ 
& $\chi^2_{\rm base}/{\rm dof}$ 
& $\chi^2_{\rm PL}/{\rm dof}$ 
& $\Delta\chi^2$ \\
\hline
2685 & $2.08^{+0.30}_{-0.50}$ & 1182.6/1285 & 1050.3/1283 & 132.3 \\
2686 & $2.43^{+0.04}_{-0.17}$ & 1063.5/1285 & 931.1/1283  & 132.3 \\
2687 & $2.45^{+0.03}_{-0.03}$ & 1124.8/1285 & 1036.0/1283 & 88.8 \\
2689 & $2.53^{+0.04}_{-0.03}$ & 1003.5/1285 & 948.5/1283  & 55.1 \\
2690 & $2.51^{+0.08}_{-0.04}$ & 1039.6/1285 & 969.4/1283  & 70.2 \\
2691 & $2.52^{+0.02}_{-0.05}$ & 992.5/1285  & 958.5/1283  & 34.0 \\
2692 & $2.59^{+0.05}_{-0.05}$ & 1005.4/1285 & 933.3/1283  & 72.2 \\
2693 & $2.61^{+0.02}_{-0.02}$ & 1053.1/1285 & 957.6/1283  & 95.5 \\
2694 & $2.43^{+0.03}_{-0.03}$ & 1127.8/1285 & 989.1/1283  & 138.7 \\
\hline
\end{tabular}
\end{table*}

An additional high-energy component beyond the thermal Comptonization continuum has been reported at different stages of the 2023 outburst of Swift J1727.8-1613 \citep{2024ApJPengJQ,Mereminskiy2024,2024A&A_Bouchet,Caojia2025,Chand2026,Liu2026}. Based on Insight-HXMT data, \citet{2025arXiv_He} found that the 2--$100 \, {\rm keV}$ spectra of Swift J1727.8-1613 during the flare state were well described by the model \texttt{constant*tbabs*(thcomp*diskbb+relxillCp)}, which accounts for thermal Comptonization of disk seed photons and relativistic disk reflection. However, this energy range does not provide sufficient coverage above the thermal Comptonization continuum to constrain the high-energy emission. Using INTEGRAL observations extending up to $\sim500\,{\rm keV}$, \citet{2024A&A_Bouchet} detected a high-energy tail throughout the hard intermediate state (HIMS); a similar component was also identified in the Insight-HXMT spectra \citep[][]{Caojia2025}. More recently, \citet{Chand2026} showed that the excess above $\sim100\,{\rm keV}$ in a single HIMS observation could be reproduced by assuming a hybrid thermal/non-thermal electron distribution. Motivated by these results, we used nine joint Insight-HXMT and INTEGRAL spectra spanning 2--500 keV to investigate the spectral and physical evolution of the high-energy excess during the flare state.

Following the spectral fitting procedure described in Section \ref{sec:spectral_analysis}, we first applied the fitting model \texttt{constant*tbabs*(thcomp*diskbb+relxillCp)} to the joint spectra. The fit to revolution 2685 is shown as a representative example in the left panel of Figure~\ref{fig:excess}. Although the model adequately describes the spectra below $\sim100\,{\rm keV}$, clear systematic positive residuals are present above $\sim150\,{\rm keV}$. We therefore added a phenomenological power-law component to account for the high-energy excess, yielding a model \texttt{constant*tbabs*}\allowbreak\texttt{(thcomp*diskbb+}\allowbreak\texttt{relxillCp}\allowbreak\texttt{+powerlaw)}. As shown in the right panel of Figure~\ref{fig:excess}, the additional component removes the residuals above $\sim150\,{\rm keV}$ and improves the fit from $\chi^2/{\rm dof}=1182.6/1285$ to $1050.3/1283$. To estimate the F-test probability for adding the power-law component, we performed 1000 simulations using the \texttt{simftest} script in \texttt{XSPEC} \footnote{\url{https://heasarc.gsfc.nasa.gov/docs/software/xspec/manual/node126.html}}. The resulting probability is $p<10^{-3}$, indicating that the power-law component is statistically required.

The same model was then applied to the other eight spectra. Similar high-energy residuals are generally present throughout the flare state. We found that adding the power-law component improves the fits by $\Delta\chi^2=34.0$--$138.7$ for two additional free parameters. Throughout the flare state, the photon index of the additional power-law component remains within the range $\Gamma_{\rm PL}=2.1$--2.6. The best-fit photon indices and fit statistics are summarized in Table~\ref{tab:powerlaw}.

\subsection{Hybrid Comptonization Model}

\begin{figure*}[htbp]
    \centering
    \includegraphics[scale=0.48]{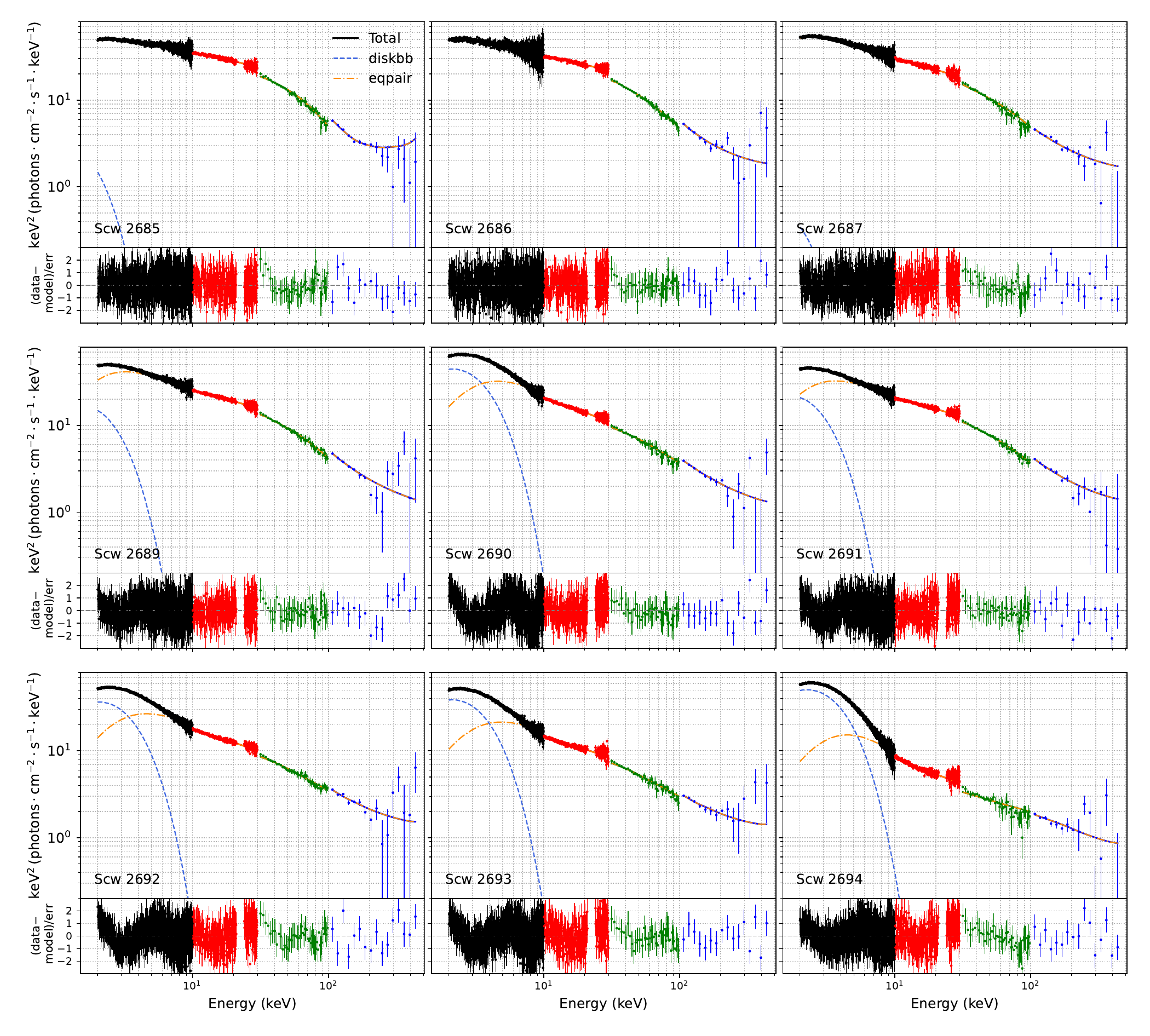}
    \caption{Broadband spectral fits with the \texttt{constant*babs*(diskbb+eqpair)} model for nine INTEGRAL revolutions during the flare state. The black, red, green, and blue points represent the Insight-HXMT/LE, ME, HE, and INTEGRAL/ISGRI data, respectively. The solid curves show the total model, while the dashed and dash-dotted curves show the \texttt{diskbb} and \texttt{eqpair} components, respectively. The lower panels show the residuals in units of the statistical error.}
    \label{fig:eqpair}
\end{figure*}

The analysis above shows that the high-energy excess is present throughout the flare state. Such an excess may arise from an additional Comptonizing region, Comptonization by a hybrid thermal/non-thermal electron population, or emission associated with the jet \citep[e.g.,][]{Droulans2010,McConnell2002ApJ,Jourdain2012jet,Bassi2020}. In particular, the high-energy emission of Cyg X-1, GX 339-4, and GRS 1716-249 has been successfully modeled with the hybrid electron Comptonization model \texttt{eqpair} \citep{McConnell2002ApJ,DelSanto2008,Bassi2020}, which self-consistently calculates the equilibrium electron distribution and the resulting spectrum \citep{Coppi1999}.

\begin{table*}
\centering
\caption{Best-fit parameters of the \texttt{constant*tbabs*(diskbb+eqpair)} model for INTEGRAL revolutions 2685--2694 during the flare state of Swift J1727.8-1613. Uncertainties are quoted at the 90\% confidence level for a single parameter. }
\label{tab:eqpair}
\setlength{\tabcolsep}{4pt}
\begin{tabular}{lccccccc}
\hline\hline
Revolution & $l_{\rm h}/l_{\rm s}$ & $kT_{\rm bb}$ (eV) & $l_{\rm nth}/l_{\rm h}$ & $\tau_{\rm p}$ & $\Gamma_{\rm inj}$ & $\Omega/2\pi$ & $\chi^2$/dof \\
\hline
2685 & $1.617^{+0.051}_{-0.071}$ & $350^{+32}_{-41}$ & $0.242^{+0.019}_{-0.018}$ & $2.510^{+0.117}_{-0.050}$ & $1.48^{+0.51}_{-0.48}$ & $0.028^{+0.020}_{-0.024}$ & $1072.6/1284$ (0.835) \\
2686 & $1.399^{+0.022}_{-0.046}$ & $342^{+50}_{-10}$ & $0.607^{+0.090}_{-0.082}$ & $2.958^{+0.111}_{-0.067}$ & $3.10^{+0.36}_{-0.16}$ & $0.059^{+0.032}_{-0.045}$ & $1178.8/1284$ (0.918) \\
2687 & $1.134^{+0.018}_{-0.010}$ & $398^{+82}_{-11}$ & $0.774^{+0.106}_{-0.040}$ & $2.775^{+0.091}_{-0.053}$ & $3.29^{+0.20}_{-0.12}$ & $0.051^{+0.019}_{-0.011}$ & $1053.7/1284$ (0.821) \\
2689 & $1.095^{+0.006}_{-0.020}$ & $535^{+21}_{-26}$ & $>0.982$ & $2.716^{+0.064}_{-0.051}$ & $3.68^{+0.09}_{-0.15}$ & $0.055^{+0.016}_{-0.016}$ & $1027.7/1284$ (0.800) \\
2690 & $0.909^{+0.015}_{-0.010}$ & $806^{+6}_{-3}$ & $>0.982$ & $2.104^{+0.024}_{-0.028}$ & $3.39^{+0.08}_{-0.05}$ & $0.135^{+0.031}_{-0.015}$ & $1272.2/1284$ (0.991) \\
2691 & $1.105^{+0.008}_{-0.030}$ & $613^{+14}_{-14}$ & $>0.987$ & $2.727^{+0.046}_{-0.076}$ & $3.46^{+0.14}_{-0.09}$ & $0.069^{+0.017}_{-0.014}$ & $1072.5/1284$ (0.835) \\
2692 & $1.022^{+0.041}_{-0.023}$ & $772^{+8}_{-7}$ & $>0.986$ & $2.383^{+0.034}_{-0.014}$ & $3.12^{+0.06}_{-0.06}$ & $0.154^{+0.026}_{-0.022}$ & $1169.5/1284$ (0.911) \\
2693 & $1.082^{+0.029}_{-0.048}$ & $805^{+2}_{-7}$ & $>0.969$ & $2.440^{+0.016}_{-0.011}$ & $3.00^{+0.09}_{-0.04}$ & $0.177^{+0.008}_{-0.035}$ & $1304.6/1284$ (1.016) \\
2694 & $0.746^{+0.022}_{-0.025}$ & $869^{+5}_{-2}$ & $>0.991$ & $1.052^{+0.023}_{-0.018}$ & $2.93^{+0.05}_{-0.01}$ & $0.324^{+0.070}_{-0.042}$ & $1197.0/1284$ (0.932) \\
\hline
\end{tabular}
\end{table*}

In \texttt{eqpair}, the luminosity $L$ is expressed in terms of the dimensionless compactness parameter, $l=L\sigma_{\rm T}/(Rm_{\rm e}c^3)$, where $R$ is the characteristic radius of the emitting region. The soft compactness $l_{\rm s}$ is proportional to the luminosity of the thermal disk photons entering the Comptonizing region. The hard compactness $l_{\rm h}=l_{\rm th}+l_{\rm nth}$ represents the total power supplied to the electrons in the plasma, where $l_{\rm th}$ and $l_{\rm nth}$ describe thermal heating and non-thermal acceleration of the electrons, respectively. Therefore, the ratios $l_{\rm h}/l_{\rm s}$ and $l_{\rm nth}/l_{\rm h}$ mainly determine the spectral shape. The continuous acceleration of electrons is described by a power-law distribution $Q(\gamma)\propto\gamma^{-\Gamma_{\rm inj}}$. We fixed the lower and upper bounds of the injected distribution at $\gamma_{\rm min}=1.3$ and $\gamma_{\rm max}=1000$, respectively, following commonly adopted \texttt{eqpair} settings in previous studies of black hole X-ray binaries \citep[e.g.,][]{Coppi1999,DelSanto2008,DelSanto2016,Bassi2020}. The plasma is further characterized by the seed-photon temperature $kT_{\rm bb}$, and the Thomson scattering optical depth $\tau_{\rm p}$ of the background plasma. For a given set of input parameters, the equilibrium electron/positron distribution, electron temperature, total optical depth, and resulting spectrum are calculated self-consistently \citep{Coppi1999}.

To account for disk photons that escape the Comptonizing region without being scattered, \citet{Coppi1999} suggested adding a separate \texttt{diskbb} component to \texttt{eqpair}. We therefore fitted the broadband spectra during the flare state with the model \texttt{constant*tbabs* (diskbb+eqpair)}. We adopted the same disk inclination of $40^\circ$ used in the fitting described in Section~\ref{sec:spectral_analysis}. The disk blackbody temperature of \texttt{diskbb} was tied to $kT_{\rm bb}$ in \texttt{eqpair}, while its normalization was allowed to vary. Following previous applications of \texttt{eqpair} to black hole X-ray binaries, we fixed the soft compactness at $l_{\rm s}=10$ \citep[e.g.,][]{DelSanto2008,Bassi2020}, since the spectral shape in this regime depends mainly on the compactness ratios rather than the exact value of $l_{\rm s}$. $R=10^7\,{\rm cm}$ was adopted in the fitting as a typical value. The reflection parameter, $\Omega/2\pi$, was free to fit. Previous reflection analyses of Swift J1727.8-1613 found a highly ionized accretion disk during the flare state, with ${\rm log} \, \xi \sim3.7-4.3$ \citep[e.g.,][]{Xusaien2025ApJ,Chand2026,Majumder2026}. We therefore fixed $\xi=5000\,{\rm erg\,cm\,s^{-1}}$, the upper limit allowed by \texttt{eqpair}. The elemental abundances were fixed at the solar values. The reflector temperature was fixed at $10^6 \,{\rm K}$, the radial emissivity index was fixed at $\beta=3$, and the inner and outer radii of the reflecting disk were fixed at $10 \, R_{\rm g}$ and $1000 \, R_{\rm g}$, respectively.

The free parameters in our fits were $l_{\rm h}/l_{\rm s}$, $l_{\rm nth}/l_{\rm h}$, $kT_{\rm bb}$, $\tau_{\rm p}$, $\Gamma_{\rm inj}$, the reflection parameter $\Omega/2\pi$, and the model normalization. The best-fitting spectra are shown in Figure~\ref{fig:eqpair} and the corresponding parameters are summarized in Table~\ref{tab:eqpair}. All nine spectra are well described by the hybrid Comptonization model, with reduced $\chi^2$ values ranging from 0.80 to 1.02.

\begin{figure*}[htbp]
    \centering
    \includegraphics[scale=0.63]{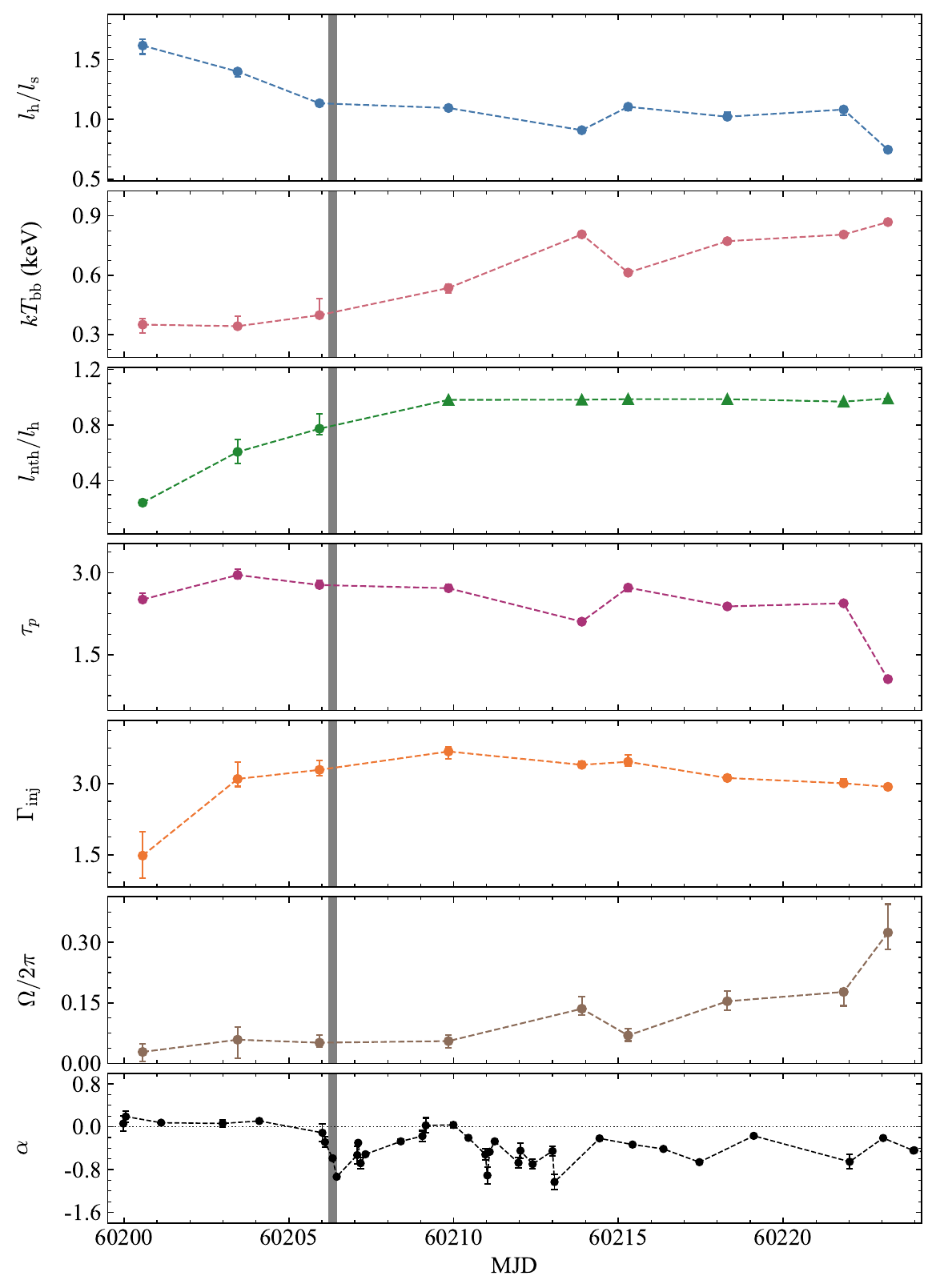}
    \caption{Evolution of the best-fitting \texttt{eqpair} parameters during the flare state. The upward triangles indicate the 90\% confidence lower limits on $l_{\rm nth}/l_{\rm h}$. The radio spectral indices are taken from \citet{Hughes2025}. The gray shaded region marks the inferred ejection interval of the three transient jet knots reported by \citet{Wood2025}.}
    \label{fig:evolution}
\end{figure*}

The evolution of the best-fitting parameters is shown in Figure~\ref{fig:evolution}. During the flare state, $l_{\rm h}/l_{\rm s}$ shows an overall decrease from 1.62 to 0.75, while $kT_{\rm bb}$ increases from approximately 0.35 to $0.87 \, {\rm keV}$. These trends are consistent with the spectral softening expected as the source evolves through the HIMS from the hard state toward the soft state. The reflection parameter $\Omega/2\pi$ shows an overall increase from about 0.03 to 0.32, which is consistent with the inward evolution of the inner disk boundary and/or a more compact coronal geometry \citep{2025arXiv_He,Liao2025}. The Thomson scattering depth remains within $\tau_{\rm p}\simeq2.1-3.0$ over most of the period, decreasing only in revolution 2694. Most notably, $l_{\rm nth}/l_{\rm h}$ increases from 0.24 in revolution 2685 to 0.77 in revolution 2687. From revolution 2689 onward, the best-fitting value reaches the upper boundary, with 90\% confidence lower limits ranging from 0.969 to 0.991, indicating that the energy supply to the electrons changes from predominantly thermal heating to non-thermal acceleration and remains dominated by non-thermal acceleration thereafter. Meanwhile, $\Gamma_{\rm inj}$ increases from 1.48 in revolution 2685 to 3.10 in revolution 2686 and remains within approximately 2.93--3.68 thereafter, showing that the injected electron distribution becomes substantially steeper after revolution 2685.

\section{Discussion and Conclusion} \label{sec:4}

Our broadband spectral analysis, covering 2--$500\, {\rm keV}$, shows that the high-energy excess is present throughout the flare state of Swift J1727.8-1613. A model consisting of disk emission, thermal Comptonization, and relativistic disk reflection leaves clear positive residuals above $\sim150\,{\rm keV}$. Adding a phenomenological power-law component to this model significantly improves the fits, yielding photon indices of $\Gamma_{\rm PL} \simeq 2.1$--$2.6$, broadly consistent with those reported in previous studies \citep[e.g.,][]{2024A&A_Bouchet,Caojia2025}. To explore the physical origin of the high-energy excess, we then fitted the broadband spectra with the hybrid Comptonization model \texttt{eqpair}. We found that all nine broadband spectra spanning the flare state can be well described by \texttt{eqpair}, without requiring an additional component to account for the high-energy excess. Within this model, $l_{\rm nth}/l_{\rm h}$ increases from 0.24 (revolution 2685) to nearly unity (revolution 2689) and remains at this level thereafter, indicating that the power supplied to the electrons changes from thermal heating to predominantly non-thermal acceleration. In contrast, the narrow range of photon indices obtained for the phenomenological power-law component indicates that the overall shape of the high-energy excess does not change substantially during the same period. Thus, although a phenomenological power-law provides a useful description of the excess, it may obscure the underlying changes of the emitting plasma. This highlights the importance of using physical models to investigate the high-energy excess when sufficient broadband data are available.

The fitting results with \texttt{eqpair} suggest that the high-energy excess during the flare state in Swift J1727.8-1613 may arise from Comptonization by non-thermal electrons in a hybrid plasma. Similar interpretations have been proposed for the hard-state and early-HIMS spectra of Swift J1727.8-1613 \citep[e.g.,][]{Liu2026,Chand2026}. However, \citet{Liu2026} found that their \texttt{eqpair} fits to the earlier rising hard-state spectra required extreme compactness and injection-index values, and therefore adopted a more self-consistent treatment of Comptonization, disk reflection, and relativistic effects. This may be because the reflection spectrum in the rising hard state is more complex \citep[e.g.,][]{2024ApJPengJQ,Peng2025}, whereas \texttt{eqpair} provides only an approximate, non-relativistic treatment of disk reflection. In contrast, the reflection component during the flare state is weak \citep[e.g.,][]{Xusaien2025ApJ,Caojia2025}, and the approximate reflection treatment in \texttt{eqpair} is therefore expected to have a smaller influence on our spectral fits. Contemporaneous radio observations revealed bright radio flares during the flare state, and three transient jet knots were identified through time-dependent modeling of the VLBI data \citep{Hughes2025,Wood2025}. Transient ejecta are generally associated with optically thin synchrotron emission \citep{Fender2004,Fender2009}. More generally, jet emission has been observed or modeled to extend into the soft gamma-ray band through synchrotron and synchrotron self-Compton processes \citep{Laurent2011,Jourdain2012jet,Zdziarski2017}. A jet contribution to the high-energy excess of Swift J1727.8-1613 has been suggested from spectral modeling by \citet{Caojia2025}. More direct evidence for a jet contribution was provided by \citet{2024A&A_Bouchet}, which detected strong polarization above 210 keV during the HIMS and argued that the measured polarization degree is difficult to explain through Comptonization alone. However, the polarization angle was not aligned with the projected direction of the compact radio jet, making a direct association of the polarized emission with the radio jet less straightforward \citep{2024A&A_Bouchet}. \citet{2024A&A_Bouchet} suggested that this misalignment may result from a complex geometry near the jet base or a non-uniform pitch-angle distribution of the emitting electrons. On the other hand, the high-energy excess had already been detected during the hard state and early HIMS \citep{Mereminskiy2024,2024ApJPengJQ,Liu2026,Chand2026}, before the transient jet knots reported by \citet{Wood2025}, indicating that these transient ejecta cannot be the sole origin of the excess throughout the outburst. In our analysis, broadband spectra during the flare state can be reproduced with a single hybrid Comptonization component, without introducing a separate high-energy continuum. This indicates that non-thermal Comptonization provides a sufficient and consistent description of the high-energy excess throughout the flare state, although the possible contribution from the compact jet or jet base cannot be excluded \citep[e.g.,][]{Mereminskiy2024,2024ApJPengJQ,Caojia2025,Peng2025}.

The overall evolution of the best-fitting \texttt{eqpair} parameters is consistent with that expected as the source evolves from the hard state toward the soft state. The most notable parameters are $l_{\rm nth}/l_{\rm h}$ and $\Gamma_{\rm inj}$, which are closely related to the high-energy excess. $l_{\rm nth}/l_{\rm h}$ increases from 0.24 in revolution 2685 to nearly unity in revolution 2689, and remains at this level thereafter, indicating that the power supplied to the electrons changes from predominantly thermal heating to predominantly non-thermal acceleration. This increase is not the consequence of a decrease in $l_{\rm h}$. The ratio $l_{\rm nth}/l_{\rm s}=(l_{\rm nth}/l_{\rm h})(l_{\rm h}/l_{\rm s})$ increases from approximately 0.39 in revolution 2685 to $\gtrsim1.08$ in revolution 2689, demonstrating that, relative to the fixed soft compactness, the inferred non-thermal power itself increases by at least a factor of approximately 2.7. Meanwhile, $\Gamma_{\rm inj}$ increases abruptly from approximately 1.5 in revolution 2685 to 3.1 in revolution 2686 and remains within 2.9--3.7 thereafter. The increase in the electron injection index does not imply weaker non-thermal acceleration; instead, it reflects a change in the distribution of the non-thermal power, with a larger fraction supplied to electrons with low and intermediate Lorentz factors rather than to the highest-energy electrons. Together, these results indicate a substantial change in the electron injection properties of the plasma during the flare state. Several physical mechanisms may account for the inferred evolution of the electron injection. Magnetic reconnection in the corona or at the jet base can dissipate magnetic energy and accelerate non-thermal electrons \citep[e.g.][]{Beloborodov2017,Sironi2020}. Non-thermal electrons may also be produced by stochastic interactions with plasma turbulence \citep[e.g.,][]{Lihui1997,Comisso2018_prl,Dermer1996ApJ}, or through first-order Fermi acceleration at collisionless shocks that may develop within transient ejecta \citep{Blandford1978,Sironi2015}. Although our results do not distinguish among these acceleration mechanisms, the fact that high-energy excess was already present before the ejection of the transient jet knots indicates that shock acceleration within the ejecta cannot be the sole origin of the high-energy excess.

The radio spectral indices reported by \citet{Hughes2025} are shown in the bottom panel of Figure~\ref{fig:evolution} for comparison. Around MJD$\sim$60206, the radio spectral index decreased rapidly from approximately zero to $\alpha\sim-1$. A flat or slightly inverted radio spectrum is characteristic of the synchrotron self-absorption emission of a compact jet \citep{Blandford1979}, whereas a negative spectral index is expected for optically thin synchrotron emission from a transient jet \citep{Fender2001,2006csxs.book..381Fender}. \citet{Wood2025} constrained the ejection times of three transient jet knots to MJD~60206.22--60206.41, as marked by the gray shaded region in Figure~\ref{fig:evolution}. This interval occurs after the abrupt increase in $\Gamma_{\rm inj}$ and between revolutions 2687 and 2689, during which $l_{\rm nth}/l_{\rm h}$ increased from 0.77 to $>0.98$. The temporal association suggests that the launching of the transient jets may be related to the changes in the non-thermal electrons. The radio spectral index also allows a comparison with the non-thermal electron distribution constrained by our fits. For an electron distribution $N(\gamma)\propto\gamma^{-p}$, optically thin synchrotron emission has a spectral index $\alpha=-(p-1)/2$. If the electrons in the ejecta were injected with a slope similar to our spectral fitting results and their distribution was not significantly modified by Compton cooling, then $p\simeq\Gamma_{\rm inj}$, which gives $\alpha\sim-1$. While efficient Compton cooling would steepen the steady-state electron distribution to $p\simeq\Gamma_{\rm inj}+1$ and yield $\alpha\sim-1.5$. Indeed, some of the radio spectral indices measured after the jet ejections fall within the range $-1.5 < \alpha < -1$. An order-of-magnitude comparison can also be made for the energy budget. Around the ejection epoch reported by \citet{Wood2025}, the non-thermal compactness derived from our fits corresponds to $L_{\rm nth}=l_{\rm nth} (Rm_{\rm e}c^3)/\sigma_{\rm T} \sim3\times10^{36}\,{\rm erg\,s^{-1}}$. The ejection times of the three jet knots span approximately 0.19 days, corresponding to $\sim1.7\times10^{40}\,{\rm erg}$ for each ejection, comparable to the minimum energy of (0.6--2)$\times10^{40}\,{\rm erg}$ estimated for individual radio flares by \citet{Hughes2025}. Although these estimates are quite crude, the agreement in radio slope and energy scale suggests a possible connection between the increase in non-thermal power indicated by our X-ray spectral analysis and the ejection of the transient jet.

\begin{acknowledgments}
\section*{Acknowledgments}

This work is supported by the Natural Science Foundation of China (NSFC) 12322307, 12273026, and 12361131579; by “the Fundamental Research Funds for the Central Universities”; Xiaomi Foundation / Xiaomi Young Talents Program; The data analysis in this paper have been done on the supercomputing system in the Supercomputing Center of Wuhan University.

\end{acknowledgments}

\bibliography{1727}{}

@ARTICLE{HXMT-Zhang2020,
       author = {{Zhang}, Shuang-Nan and {Li}, TiPei and {Lu}, FangJun and {Song}, LiMing and {Xu}, YuPeng and {Liu}, CongZhan and {Chen}, Yong and {Cao}, XueLei and {Bu}, QingCui and {Chang}, Zhi and {Chen}, Gang and {Chen}, Li and {Chen}, TianXiang and {Chen}, YiBao and {Chen}, YuPeng and {Cui}, Wei and {Cui}, WeiWei and {Deng}, JingKang and {Dong}, YongWei and {Du}, YuanYuan and {Fu}, MinXue and {Gao}, GuanHua and {Gao}, He and {Gao}, Min and {Ge}, MingYu and {Gu}, YuDong and {Guan}, Ju and {Gungor}, Can and {Guo}, ChengCheng and {Han}, DaWei and {Hu}, Wei and {Huang}, Yue and {Huo}, Jia and {Jia}, ShuMei and {Jiang}, LuHua and {Jiang}, WeiChun and {Jin}, Jing and {Jin}, YongJie and {Li}, Bing and {Li}, ChengKui and {Li}, Gang and {Li}, MaoShun and {Li}, Wei and {Li}, Xian and {Li}, XiaoBo and {Li}, XuFang and {Li}, YanGuo and {Li}, ZiJian and {Li}, ZhengWei and {Liang}, XiaoHua and {Liao}, JinYuan and {Liu}, GuoQing and {Liu}, HongWei and {Liu}, ShaoZhen and {Liu}, XiaoJing and {Liu}, Yuan and {Liu}, YiNong and {Lu}, Bo and {Lu}, XueFeng and {Luo}, Tao and {Ma}, Xiang and {Meng}, Bin and {Nang}, Yi and {Nie}, JianYin and {Ou}, Ge and {Qu}, JinLu and {Sai}, Na and {Shang}, RenCheng and {Shen}, GuoHong and {Sun}, Liang and {Tan}, Ying and {Tao}, Lian and {Tuo}, YouLi and {Wang}, Chen and {Wang}, ChunQin and {Wang}, GuoFeng and {Wang}, HuanYu and {Wang}, Juan and {Wang}, WenShuai and {Wang}, YuSa and {Wen}, XiangYang and {Wu}, BaiYang and {Wu}, BoBing and {Wu}, Mei and {Xiao}, GuangCheng and {Xiong}, ShaoLin and {Yan}, LinLi and {Yang}, JiaWei and {Yang}, Sheng and {Yang}, YanJi and {Yi}, QiBin and {Yuan}, Bin and {Zhang}, AiMei and {Zhang}, ChunLei and {Zhang}, ChengMo and {Zhang}, Fan and {Zhang}, HongMei and {Zhang}, Juan and {Zhang}, Qiang and {Zhang}, ShenYi and {Zhang}, Shu and {Zhang}, Tong and {Zhang}, WanChang and {Zhang}, Wei and {Zhang}, WenZhao and {Zhang}, Yi and {Zhang}, YiFei and {Zhang}, YongJie and {Zhang}, Yue and {Zhang}, Zhao and {Zhang}, Zhi and {Zhang}, ZiLiang and {Zhao}, HaiSheng and {Zhao}, XiaoFan and {Zheng}, ShiJie and {Zhou}, JianFeng and {Zhu}, YuXuan and {Zhu}, Yue and {Zhuang}, RenLin and {The Insight-HXMT team}},
        title = "{Overview to the Hard X-ray Modulation Telescope (Insight-HXMT) Satellite}",
      journal = {Science China Physics, Mechanics, and Astronomy},
         year = 2020,
        month = apr,
       volume = {63},
       number = {4},
          eid = {249502},
        pages = {249502},
          doi = {10.1007/s11433-019-1432-6},
archivePrefix = {arXiv},
       eprint = {1910.09613},
 primaryClass = {astro-ph.IM},
       adsurl = {https://ui.adsabs.harvard.edu/abs/2020SCPMA..6349502Z}
}

@ARTICLE{veledina2023,
       author = {{Veledina}, Alexandra and {Muleri}, Fabio and {Dov{\v{c}}iak}, Michal and {Poutanen}, Juri and {Ratheesh}, Ajay and {Capitanio}, Fiamma and {Matt}, Giorgio and {Soffitta}, Paolo and {Tennant}, Allyn F. and {Negro}, Michela and et al.},
        title = "{Discovery of X-Ray Polarization from the Black Hole Transient Swift J1727.8-1613}",
      journal = {\apjl},
         year = 2023,
        month = nov,
       volume = {958},
       number = {1},
          eid = {L16},
        pages = {L16},
          doi = {10.3847/2041-8213/ad0781},
archivePrefix = {arXiv},
       eprint = {2309.15928},
 primaryClass = {astro-ph.HE},
       adsurl = {https://ui.adsabs.harvard.edu/abs/2023ApJ...958L..16V}
}

@ARTICLE{li2026,
       author = {{Li}, Ze-Xi and {Zhang}, Liang and {Tao}, Lian and {Yang}, Zi-Han and {Zhao}, Qing-Chang and {Zhao}, Shu-Jie and {Ma}, Rui-Can and {Yang}, Zi-Xu and {Li}, Pan-Ping and {Ma}, Xiang and et al.},
        title = "{Characteristics of the high-frequency humps in the black hole X-ray binary Swift J1727.8─1613}",
      journal = {\aap},
         year = 2026,
        month = feb,
       volume = {707},
          eid = {A33},
        pages = {A33},
          doi = {10.1051/0004-6361/202555276},
archivePrefix = {arXiv},
       eprint = {2601.03696},
 primaryClass = {astro-ph.HE},
       adsurl = {https://ui.adsabs.harvard.edu/abs/2026A&A...707A..33L}
}

@ARTICLE{rawat2025,
       author = {{Rawat}, Divya and {M{\'e}ndez}, Mariano and {Garc{\'\i}a}, Federico and {Maggi}, Pierre},
        title = "{Evolution of the Comptonizing medium of the black-hole candidate Swift J1727.8─1613 along the hard to hard-intermediate state transition using NICER}",
      journal = {\aap},
         year = 2025,
        month = may,
       volume = {697},
          eid = {A229},
        pages = {A229},
          doi = {10.1051/0004-6361/202453538},
archivePrefix = {arXiv},
       eprint = {2504.06705},
 primaryClass = {astro-ph.HE},
       adsurl = {https://ui.adsabs.harvard.edu/abs/2025A&A...697A.229R}
}

@ARTICLE{jin2025,
       author = {{Jin}, Pei and {M{\'e}ndez}, Mariano and {Garc{\'\i}a}, Federico and {Altamirano}, Diego and {Zhang}, Guobao and {Rout}, Sandeep K.},
        title = "{Timing analysis of the black hole candidate Swift J1727.8─1613: Detection of a dip-like feature in the high-energy cross spectrum}",
      journal = {\aap},
         year = 2025,
        month = jun,
       volume = {699},
          eid = {A9},
        pages = {A9},
          doi = {10.1051/0004-6361/202554353},
archivePrefix = {arXiv},
       eprint = {2504.20717},
 primaryClass = {astro-ph.HE},
       adsurl = {https://ui.adsabs.harvard.edu/abs/2025A&A...699A...9J}
}

@ARTICLE{cangemi2021,
       author = {{Cangemi}, F. and {Beuchert}, T. and {Siegert}, T. and {Rodriguez}, J. and {Grinberg}, V. and {Belmont}, R. and {Gouiff{\`e}s}, C. and {Kreykenbohm}, I. and {Laurent}, P. and {Pottschmidt}, K. and {Wilms}, J.},
        title = "{Potential origin of the state-dependent high-energy tail in the black hole microquasar Cygnus X-1 as seen with INTEGRAL}",
      journal = {\aap},
         year = 2021,
        month = jun,
       volume = {650},
          eid = {A93},
        pages = {A93},
          doi = {10.1051/0004-6361/202038604},
archivePrefix = {arXiv},
       eprint = {2102.04773},
 primaryClass = {astro-ph.HE},
       adsurl = {https://ui.adsabs.harvard.edu/abs/2021A&A...650A..93C}
}

@ARTICLE{carotenuto2021,
       author = {{Carotenuto}, F. and {Corbel}, S. and {Tremou}, E. and {Russell}, T.~D. and {Tzioumis}, A. and {Fender}, R.~P. and {Woudt}, P.~A. and {Motta}, S.~E. and {Miller-Jones}, J.~C.~A. and {Chauhan}, J. and {Tetarenko}, A.~J. and {Sivakoff}, G.~R. and {Heywood}, I. and {Horesh}, A. and {van der Horst}, A.~J. and {Koerding}, E. and {Mooley}, K.~P.},
        title = "{The black hole transient MAXI J1348-630: evolution of the compact and transient jets during its 2019/2020 outburst}",
      journal = {\mnras},
         year = 2021,
        month = jun,
       volume = {504},
       number = {1},
        pages = {444-468},
          doi = {10.1093/mnras/stab864},
archivePrefix = {arXiv},
       eprint = {2103.12190},
 primaryClass = {astro-ph.HE},
       adsurl = {https://ui.adsabs.harvard.edu/abs/2021MNRAS.504..444C}
}

@ARTICLE{you2024,
       author = {{You}, Bei and {Yang}, Shuai-kang and {Yan}, Zhen and {Cao}, Xinwu and {Zdziarski}, Andrzej A.},
        title = "{The Delayed Radio Emission in the Black Hole X-Ray Binary MAXI J1348-630}",
      journal = {\apjl},
         year = 2024,
        month = jul,
       volume = {969},
       number = {2},
          eid = {L33},
        pages = {L33},
          doi = {10.3847/2041-8213/ad5b50},
archivePrefix = {arXiv},
       eprint = {2406.15994},
 primaryClass = {astro-ph.HE},
       adsurl = {https://ui.adsabs.harvard.edu/abs/2024ApJ...969L..33Y}
}

@ARTICLE{you2023,
       author = {{You}, Bei and {Cao}, Xinwu and {Yan}, Zhen and {Hameury}, Jean-Marie and {Czerny}, Bozena and {Wu}, Yue and {Xia}, Tianyu and {Sikora}, Marek and {Zhang}, Shuang-Nan and {Du}, Pu and {Zycki}, Piotr T.},
        title = "{Observations of a black hole x-ray binary indicate formation of a magnetically arrested disk}",
      journal = {Science},
         year = 2023,
        month = sep,
       volume = {381},
       number = {6661},
        pages = {961-964},
          doi = {10.1126/science.abo4504},
archivePrefix = {arXiv},
       eprint = {2309.00200},
 primaryClass = {astro-ph.HE},
       adsurl = {https://ui.adsabs.harvard.edu/abs/2023Sci...381..961Y}
}

@ARTICLE{bright2020,
       author = {{Bright}, J.~S. and {Fender}, R.~P. and {Motta}, S.~E. and {Williams}, D.~R.~A. and {Moldon}, J. and {Plotkin}, R.~M. and {Miller-Jones}, J.~C.~A. and {Heywood}, I. and {Tremou}, E. and {Beswick}, R. and {Sivakoff}, G.~R. and {Corbel}, S. and {Buckley}, D.~A.~H. and {Homan}, J. and {Gallo}, E. and {Tetarenko}, A.~J. and {Russell}, T.~D. and {Green}, D.~A. and {Titterington}, D. and {Woudt}, P.~A. and {Armstrong}, R.~P. and {Groot}, P.~J. and {Horesh}, A. and {van der Horst}, A.~J. and {K{\"o}rding}, E.~G. and {McBride}, V.~A. and {Rowlinson}, A. and {Wijers}, R.~A.~M.~J.},
        title = "{An extremely powerful long-lived superluminal ejection from the black hole MAXI J1820+070}",
      journal = {Nature Astronomy},
         year = 2020,
        month = mar,
       volume = {4},
        pages = {697-703},
          doi = {10.1038/s41550-020-1023-5},
archivePrefix = {arXiv},
       eprint = {2003.01083},
 primaryClass = {astro-ph.HE},
       adsurl = {https://ui.adsabs.harvard.edu/abs/2020NatAs...4..697B}
}

@ARTICLE{corbel2013,
       author = {{Corbel}, S. and {Aussel}, H. and {Broderick}, J.~W. and {Chanial}, P. and {Coriat}, M. and {Maury}, A.~J. and {Buxton}, M.~M. and {Tomsick}, J.~A. and {Tzioumis}, A.~K. and {Markoff}, S. and {Rodriguez}, J. and {Bailyn}, C.~D. and {Brocksopp}, C. and {Fender}, R.~P. and {Petrucci}, P.~O. and {Cadolle-Bel}, M. and {Calvelo}, D. and {Harvey-Smith}, L.},
        title = "{Formation of the compact jets in the black hole GX 339-4.}",
      journal = {\mnras},
         year = 2013,
        month = apr,
       volume = {431},
        pages = {L107-L111},
          doi = {10.1093/mnrasl/slt018},
archivePrefix = {arXiv},
       eprint = {1303.2551},
 primaryClass = {astro-ph.HE},
       adsurl = {https://ui.adsabs.harvard.edu/abs/2013MNRAS.431L.107C}
}

@ARTICLE{du2026,
       author = {{Du}, Dizhan and {You}, Bei and {Yan}, Zhen and {Ma}, Yuao and {Cao}, Xinwu},
        title = "{Radio-X-ray Time Lags in GX 339-4: Probing Magnetic Field Transport in Black Hole Accretion}",
      journal = {arXiv e-prints},
         year = 2026,
        month = may,
          eid = {arXiv:2605.19473},
        pages = {arXiv:2605.19473},
          doi = {10.48550/arXiv.2605.19473},
archivePrefix = {arXiv},
       eprint = {2605.19473},
 primaryClass = {astro-ph.HE},
       adsurl = {https://ui.adsabs.harvard.edu/abs/2026arXiv260519473D}
}

@ARTICLE{yang2026,
       author = {{Yang}, Shuai-Kang and {You}, Bei and {Bollemeijer}, Niek and {Uttley}, Phil and {Tetarenko}, A.~J. and {Zdziarski}, Andrzej A. and {Chen}, Liang and {Casella}, P. and {Paice}, J.~A. and {Bai}, Yang and {Xu}, Sai-En},
        title = "{Covariance Spectrum of MAXI J1820+070: On the Nature of the Comptonizing Flow}",
      journal = {\apj},
         year = 2026,
        month = mar,
       volume = {1000},
       number = {1},
          eid = {20},
        pages = {20},
          doi = {10.3847/1538-4357/ae4724},
archivePrefix = {arXiv},
       eprint = {2511.17285},
 primaryClass = {astro-ph.HE},
       adsurl = {https://ui.adsabs.harvard.edu/abs/2026ApJ..1000...20Y}
}

@ARTICLE{veledina2011,
       author = {{Veledina}, Alexandra and {Vurm}, Indrek and {Poutanen}, Juri},
        title = "{A self-consistent hybrid Comptonization model for broad-band spectra of accreting supermassive black holes}",
      journal = {\mnras},
         year = 2011,
        month = jul,
       volume = {414},
       number = {4},
        pages = {3330-3343},
          doi = {10.1111/j.1365-2966.2011.18635.x},
archivePrefix = {arXiv},
       eprint = {1012.0439},
 primaryClass = {astro-ph.HE},
       adsurl = {https://ui.adsabs.harvard.edu/abs/2011MNRAS.414.3330V}
}

@ARTICLE{yu2026,
       author = {{Yu}, Wei and {Allak}, Sinan and {Yang}, Zi-Xu and {Fan}, Xiao and {Santangelo}, Andrea and {Xie}, Tian-hao},
        title = "{Stable X-ray reverberation lags in the black hole X-ray binary Swift J1727.8─1613}",
      journal = {\aap},
         year = 2026,
        month = aug,
       volume = {712},
          eid = {A75},
        pages = {A75},
          doi = {10.1051/0004-6361/202661022},
       adsurl = {https://ui.adsabs.harvard.edu/abs/2026A&A...712A..75Y}
}

@ARTICLE{ma2026mnras,
       author = {{Ma}, Ruican and {Vincentelli}, Federico and {Altamirano}, Diego and {Veledina}, Alexandra and {Gandhi}, Poshak and {Casella}, Piergiorgio and {Shahbaz}, Tariq and {Woahene-Demehin}, Sian},
        title = "{Energy-dependent Optical/Near-infrared and X-ray Correlations in Swift J1727.8-1613}",
      journal = {\mnras},
         year = 2026,
        month = aug,
          doi = {10.1093/mnras/stag1460},
       adsurl = {https://ui.adsabs.harvard.edu/abs/2026MNRAS.tmp.1366M}
}

@ARTICLE{nitindala2026,
       author = {{Nitindala}, Anagha P. and {Veledina}, Alexandra and {Kravtsov}, Vadim and {Berdyugin}, Andrei V. and {D{\'\i}az Teodori}, Mar{\'\i}a Alejandra and {Piirola}, Vilppu and {Sakanoi}, Takeshi and {Kagitani}, Masato and {Berdyugina}, Svetlana V. and {Poutanen}, Juri},
        title = "{Optical polarimetry of the accreting black hole X-ray binary Swift J1727.8{\ensuremath{-}}1613 over the state transition and radio ejections}",
      journal = {\aap},
         year = 2026,
        month = may,
       volume = {709},
          eid = {A184},
        pages = {A184},
          doi = {10.1051/0004-6361/202558488},
archivePrefix = {arXiv},
       eprint = {2512.08716},
 primaryClass = {astro-ph.HE},
       adsurl = {https://ui.adsabs.harvard.edu/abs/2026A&A...709A.184N}
}

@ARTICLE{nandi2024,
       author = {{Nandi}, Anuj and {Das}, Santabrata and {Majumder}, Seshadri and {Katoch}, Tilak and {Antia}, H.~M. and {Shah}, Parag},
        title = "{Discovery of evolving low-frequency QPOs in hard X-rays ( 100 keV) observed in black hole Swift J1727.8-1613 with AstroSat}",
      journal = {\mnras},
         year = 2024,
        month = jun,
       volume = {531},
       number = {1},
        pages = {1149-1157},
          doi = {10.1093/mnras/stae1208},
archivePrefix = {arXiv},
       eprint = {2404.17160},
 primaryClass = {astro-ph.HE},
       adsurl = {https://ui.adsabs.harvard.edu/abs/2024MNRAS.531.1149N}
}

@ARTICLE{ma2026,
       author = {{Ma}, Xiang and {Shui}, Qing-Cang and {Ge}, Ming-Yu and {Zhang}, Liang and {Qu}, Jin-Lu and {Zhang}, Shuang-Nan and {Tao}, Lian and {Song}, Li-Ming and {Zhang}, Shu and {Feng}, Hua and et al.},
        title = "{Search for the Highest-energy Quasiperiodic Oscillation in the Black Hole X-Ray Binary Candidate Swift J1727.8{\ensuremath{-}}1613}",
      journal = {\apj},
         year = 2026,
        month = may,
       volume = {1002},
       number = {2},
          eid = {180},
        pages = {180},
          doi = {10.3847/1538-4357/ae5d34},
archivePrefix = {arXiv},
       eprint = {2605.18050},
 primaryClass = {astro-ph.HE},
       adsurl = {https://ui.adsabs.harvard.edu/abs/2026ApJ..1002..180M}
}

@ARTICLE{vincentelli2025,
       author = {{Vincentelli}, F.~M. and {Shahbaz}, T. and {Casella}, P. and {Dhillon}, V.~S. and {Paice}, J. and {Altamirano}, D. and {Segura}, N. Castro and {Fender}, R. and {Gandhi}, P. and {Littlefair}, S. and et al.},
        title = "{Sub-second optical/near-infrared quasi-periodic oscillations from the black hole X-ray transient Swift J1727.8─1613}",
      journal = {\mnras},
         year = 2025,
        month = may,
       volume = {539},
       number = {3},
        pages = {2347-2361},
          doi = {10.1093/mnras/staf600},
archivePrefix = {arXiv},
       eprint = {2503.20862},
 primaryClass = {astro-ph.HE},
       adsurl = {https://ui.adsabs.harvard.edu/abs/2025MNRAS.539.2347V}
}

@ARTICLE{chatterjee2024,
       author = {{Chatterjee}, Kaushik and {Mondal}, Santanu and {Singh}, Chandra B. and {Sugizaki}, Mutsumi},
        title = "{Insight-HXMT View of the Black Hole Candidate Swift J1727.8─1613 during Its Outburst in 2023}",
      journal = {\apj},
         year = 2024,
        month = dec,
       volume = {977},
       number = {2},
          eid = {148},
        pages = {148},
          doi = {10.3847/1538-4357/ad8dc4},
archivePrefix = {arXiv},
       eprint = {2405.01498},
 primaryClass = {astro-ph.HE},
       adsurl = {https://ui.adsabs.harvard.edu/abs/2024ApJ...977..148C}
}

@ARTICLE{cao2025,
       author = {{Cao}, Hongmin and {Yang}, Jun and {Frey}, S{\'a}ndor and {Wood}, Callan M. and {Miller-Jones}, James C.~A. and {Gab{\'a}nyi}, Krisztina {\'E}. and {Migliori}, Giulia and {Giroletti}, Marcello and {Cui}, Lang and {An}, Tao and et al.},
        title = "{An Ejection Event Captured by Very Long Baseline Interferometry during the Outburst of Swift J1727.8─1613}",
      journal = {\apjl},
         year = 2025,
        month = jul,
       volume = {987},
       number = {1},
          eid = {L14},
        pages = {L14},
          doi = {10.3847/2041-8213/ade0ab},
archivePrefix = {arXiv},
       eprint = {2506.18817},
 primaryClass = {astro-ph.HE},
       adsurl = {https://ui.adsabs.harvard.edu/abs/2025ApJ...987L..14C}
}

@ARTICLE{zhu2024,
       author = {{Zhu}, Haifan and {Wang}, Wei and {Zhu}, Ziyuan},
        title = "{The Bicoherence Analysis of Type-C Quasiperiodic Oscillations in Swift J1727.8{\ensuremath{-}}1613}",
      journal = {\apj},
         year = 2024,
        month = oct,
       volume = {974},
       number = {2},
          eid = {303},
        pages = {303},
          doi = {10.3847/1538-4357/ad7587},
archivePrefix = {arXiv},
       eprint = {2408.17039},
 primaryClass = {astro-ph.HE},
       adsurl = {https://ui.adsabs.harvard.edu/abs/2024ApJ...974..303Z}
}

@ARTICLE{ingram2024,
       author = {{Ingram}, Adam and {Bollemeijer}, Niek and {Veledina}, Alexandra and {Dov{\v{c}}iak}, Michal and {Poutanen}, Juri and {Egron}, Elise and {Russell}, Thomas D. and {Trushkin}, Sergei A. and {Negro}, Michela and {Ratheesh}, Ajay and et al.},
        title = "{Tracking the X-Ray Polarization of the Black Hole Transient Swift J1727.8─1613 during a State Transition}",
      journal = {\apj},
         year = 2024,
        month = jun,
       volume = {968},
       number = {2},
          eid = {76},
        pages = {76},
          doi = {10.3847/1538-4357/ad3faf},
archivePrefix = {arXiv},
       eprint = {2311.05497},
 primaryClass = {astro-ph.HE},
       adsurl = {https://ui.adsabs.harvard.edu/abs/2024ApJ...968...76I}
}

@ARTICLE{ma2025,
       author = {{Ma}, Ruican and {Done}, Chris and {Kubota}, Aya},
        title = "{Testing the Lense─Thirring precession origin of the QPO in Swift J1727.8{\ensuremath{-}}1613}",
      journal = {\mnras},
         year = 2025,
        month = oct,
       volume = {543},
       number = {2},
        pages = {1748-1760},
          doi = {10.1093/mnras/staf1524},
archivePrefix = {arXiv},
       eprint = {2506.18857},
 primaryClass = {astro-ph.HE},
       adsurl = {https://ui.adsabs.harvard.edu/abs/2025MNRAS.543.1748M}
}

@ARTICLE{Liu2026,
       author = {{Liu}, He-Xin and {Xu}, Yan-Jun and {Yu}, Wei and {Zhang}, Shuang-Nan and {Qu}, Jin-Lu},
        title = "{The broad-band X-ray spectral properties during the rising phases of the outburst of the new black hole X-ray binary candidate Swift J1727.8─1613}",
      journal = {\mnras},
         year = 2026,
        month = jul,
       volume = {550},
       number = {1},
          eid = {stag999},
        pages = {stag999},
          doi = {10.1093/mnras/stag999},
archivePrefix = {arXiv},
       eprint = {2406.03834},
 primaryClass = {astro-ph.HE},
       adsurl = {https://ui.adsabs.harvard.edu/abs/2026MNRAS.550ag999L}
}

@ARTICLE{Cangemi2023ATel,
       author = {{Cangemi}, F. and {Bouchet}, T. and {Motta}, S.~E. and {Petrucci}, P.-O.},
        title = "{INTEGRAL detects high-energy > 200 keV emission of Swift 1727.8-1613}",
      journal = {The Astronomer's Telegram},
         year = 2023,
        month = sep,
       volume = {16238},
        pages = {1},
       adsurl = {https://ui.adsabs.harvard.edu/abs/2023ATel16238....1C}
}

@ARTICLE{Remillard2006ARAA,
       author = {{Remillard}, Ronald A. and {McClintock}, Jeffrey E.},
        title = "{X-Ray Properties of Black-Hole Binaries}",
      journal = {\araa},
         year = 2006,
        month = sep,
       volume = {44},
       number = {1},
        pages = {49-92},
          doi = {10.1146/annurev.astro.44.051905.092532},
archivePrefix = {arXiv},
       eprint = {astro-ph/0606352},
 primaryClass = {astro-ph},
       adsurl = {https://ui.adsabs.harvard.edu/abs/2006ARA&A..44...49R}
}

@ARTICLE{2023GCNPage,
       author = {{Page}, K.~L. and {Dichiara}, S. and {Gropp}, J.~D. and {Krimm}, H.~A. and {Parsotan}, T.~M. and {Williams}, M.~A. and {Neil Gehrels Swift Observatory Team}},
        title = "{GRB 230824A: Swift detection of a burst}",
      journal = {GRB Coordinates Network},
         year = 2023,
        month = aug,
       volume = {34537},
        pages = {1},
       adsurl = {https://ui.adsabs.harvard.edu/abs/2023GCN.34537....1P}
}

@ARTICLE{2024ApJPengJQ,
       author = {{Peng}, Jing-Qiang and {Zhang}, Shu and {Shui}, Qing-Cang and {Zhang}, Shuang-Nan and {Kong}, Ling-Da and {Chen}, Yu-Peng and {Wang}, Peng-Ju and {Ji}, Long and {Qu}, Jin-Lu and {Tao}, Lian and {Ge}, Ming-Yu and {Chang}, Zhi and {Li}, Jian and {Li}, Zhao-sheng and {Yu}, Zhuo-Li and {Yan}, Zhe},
        title = "{NICER, NuSTAR, and Insight-HXMT Views to the Newly Discovered Black Hole X-Ray Binary Swift J1727.8-1613}",
      journal = {\apjl},
         year = 2024,
        month = jan,
       volume = {960},
       number = {2},
          eid = {L17},
        pages = {L17},
          doi = {10.3847/2041-8213/ad17ca},
archivePrefix = {arXiv},
       eprint = {2503.01223},
 primaryClass = {astro-ph.HE},
       adsurl = {https://ui.adsabs.harvard.edu/abs/2024ApJ...960L..17P}
}

@ARTICLE{2023ATelPalmer,
       author = {{Palmer}, David M. and {Parsotan}, Tyler M.},
        title = "{Swift J1727.8-1613 reaches 7.6 Crab with strong QPO in Hard X-rays}",
      journal = {The Astronomer's Telegram},
         year = 2023,
        month = aug,
       volume = {16215},
        pages = {1},
       adsurl = {https://ui.adsabs.harvard.edu/abs/2023ATel16215....1P}
}

@ARTICLE{daixiaohang2023MNRAS,
       author = {{Dai}, Xiaohang and {Kong}, Lingda and {Bu}, Qingcui and {Santangelo}, Andrea and {Zhang}, Shu and {Ji}, Long and {Zhang}, Shuangnan and {Yorgancioglu}, Emre Seyit},
        title = "{Evolution of disc and corona in MAXI J1348-630 during the 2019 reflare: NICER and Insight-HXMT view}",
      journal = {\mnras},
         year = 2023,
        month = may,
       volume = {521},
       number = {2},
        pages = {2692-2703},
          doi = {10.1093/mnras/stad714},
archivePrefix = {arXiv},
       eprint = {2303.05290},
 primaryClass = {astro-ph.HE},
       adsurl = {https://ui.adsabs.harvard.edu/abs/2023MNRAS.521.2692D}
}

@ARTICLE{Fanxiao2026ApJ,
       author = {{Fan}, Xiao and {You}, Bei and {Du}, Dizhan and {He}, Han and {Yang}, Shuaikang},
        title = "{On the Optical Emission in the Minioutburst of the Black Hole X-Ray Binary MAXI J1348-630}",
      journal = {\apj},
         year = 2026,
        month = jan,
       volume = {997},
       number = {1},
          eid = {7},
        pages = {7},
          doi = {10.3847/1538-4357/ae2a2d},
archivePrefix = {arXiv},
       eprint = {2508.19645},
 primaryClass = {astro-ph.HE},
       adsurl = {https://ui.adsabs.harvard.edu/abs/2026ApJ...997....7F}
}

@ARTICLE{2023ATelNakajima,
       author = {{Nakajima}, M. and {Negoro}, H. and {Serino}, M. and {Mihara}, T. and {Kobayashi}, K. and {Tanaka}, M. and {Soejima}, Y. and {Kudo}, Y. and {Kawamuro}, T. and {Yamada}, S. and {Tamagawa}, T. and {Kawai}, N. and {Matsuoka}, M. and {Sakamoto}, T. and {Sugita}, S. and {Hiramatsu}, H. and {Nishikawa}, H. and {Yoshida}, A. and {Tsuboi}, Y. and {Urabe}, S. and {Nawa}, S. and {Nemoto}, N. and {Shidatsu}, M. and {Takahashi}, I. and {Niwano}, M. and {Sato}, S. and {Higuchi}, N. and {Yatsu}, Y. and {Nakahira}, S. and {Ueno}, S. and {Tomida}, H. and {Ishikawa}, M. and {Ogawa}, S. and {Kurihara}, T. and {Ueda}, Y. and {Setoguchi}, K. and {Yoshitake}, T. and {Nakatani}, Y. and {Yamauchi}, M. and {Hagiwara}, Y. and {Umeki}, Y. and {Otsuki}, Y. and {Yamaoka}, K. and {Kawakubo}, Y. and {Sugizaki}, M. and {Iwakiri}, W.},
        title = "{MAXI/GSC observations of the new X-ray transient Swift J1727.8-1613 (GRB 230824A)}",
      journal = {The Astronomer's Telegram},
         year = 2023,
        month = aug,
       volume = {16206},
        pages = {1},
       adsurl = {https://ui.adsabs.harvard.edu/abs/2023ATel16206....1N}
}

@ARTICLE{2024MNRASYu,
       author = {{Yu}, Wei and {Bu}, Qing-Cui and {Zhang}, Shuang-Nan and {Liu}, He-Xin and {Zhang}, Liang and {Ducci}, Lorenzo and {Tao}, Lian and {Santangelo}, Andrea and {Doroshenko}, Victor and {Huang}, Yue and {Yang}, Zi-Xu and {Qu}, Jin-Lu},
        title = "{Timing analysis of the newly discovered black hole candidate Swift J1727.8-1613 with Insight-HXMT}",
      journal = {\mnras},
         year = 2024,
        month = apr,
       volume = {529},
       number = {4},
        pages = {4624-4632},
          doi = {10.1093/mnras/stae835},
archivePrefix = {arXiv},
       eprint = {2403.13127},
 primaryClass = {astro-ph.HE},
       adsurl = {https://ui.adsabs.harvard.edu/abs/2024MNRAS.529.4624Y}
}

@ARTICLE{2021NatCo_You,
       author = {{You}, Bei and {Tuo}, Yuoli and {Li}, Chengzhe and {Wang}, Wei and {Zhang}, Shuang-Nan and {Zhang}, Shu and {Ge}, Mingyu and {Luo}, Chong and {Liu}, Bifang and {Yuan}, Weimin and {Dai}, Zigao and {Liu}, Jifeng and {Qiao}, Erlin and {Jin}, Chichuan and {Liu}, Zhu and {Czerny}, Bozena and {Wu}, Qingwen and {Bu}, Qingcui and {Cai}, Ce and {Cao}, Xuelei and {Chang}, Zhi and {Chen}, Gang and {Chen}, Li and {Chen}, Tianxiang and {Chen}, Yibao and {Chen}, Yong and {Chen}, Yupeng and {Cui}, Wei and {Cui}, Weiwei and {Deng}, Jingkang and {Dong}, Yongwei and {Du}, Yuanyuan and {Fu}, Minxue and {Gao}, Guanhua and {Gao}, He and {Gao}, Min and {Gu}, Yudong and {Guan}, Ju and {Guo}, Chengcheng and {Han}, Dawei and {Huang}, Yue and {Huo}, Jia and {Jia}, Shumei and {Jiang}, Luhua and {Jiang}, Weichun and {Jin}, Jing and {Jin}, Yongjie and {Kong}, Lingda and {Li}, Bing and {Li}, Chengkui and {Li}, Gang and {Li}, Maoshun and {Li}, Tipei and {Li}, Wei and {Li}, Xian and {Li}, Xiaobo and {Li}, Xufang and {Li}, Yanguo and {Li}, Zhengwei and {Liang}, Xiaohua and {Liao}, Jinyuan and {Liu}, Congzhan and {Liu}, Guoqing and {Liu}, Hongwei and {Liu}, Xiaojing and {Liu}, Yinong and {Lu}, Bo and {Lu}, Fangjun and {Lu}, Xuefeng and {Luo}, Qi and {Luo}, Tao and {Ma}, Xiang and {Meng}, Bin and {Nang}, Yi and {Nie}, Jianyin and {Ou}, Ge and {Qu}, Jinlu and {Sai}, Na and {Shang}, Rencheng and {Song}, Liming and {Song}, Xinying and {Sun}, Liang and {Tan}, Ying and {Tao}, Lian and {Wang}, Chen and {Wang}, Guofeng and {Wang}, Juan and {Wang}, Lingjun and {Wang}, Wenshuai and {Wang}, Yusa and {Wen}, Xiangyang and {Wu}, Baiyang and {Wu}, Bobing and {Wu}, Mei and {Xiao}, Guangcheng and {Xiao}, Shuo and {Xiong}, Shaolin and {Xu}, Yupeng and {Yang}, Jiawei and {Yang}, Sheng and {Yang}, Yanji and {Yi}, Qibin and {Yin}, Qianqing and {You}, Yuan and {Zhang}, Aimei and {Zhang}, Chengmo and {Zhang}, Fan and {Zhang}, Hongmei and {Zhang}, Juan and {Zhang}, Tong and {Zhang}, Wanchang and {Zhang}, Wei and {Zhang}, Wenzhao and {Zhang}, Yi and {Zhang}, Yifei and {Zhang}, Yongjie and {Zhang}, Yue and {Zhang}, Zhao and {Zhang}, Ziliang and {Zhao}, Haisheng and {Zhao}, Xiaofan and {Zheng}, Shijie and {Zhou}, Dengke and {Zhou}, Jianfeng and {Zhu}, Yuxuan and {Zhu}, Yue},
        title = "{Insight-HXMT observations of jet-like corona in a black hole X-ray binary MAXI J1820+070}",
      journal = {Nature Communications},
         year = 2021,
        month = jan,
       volume = {12},
          eid = {1025},
        pages = {1025},
          doi = {10.1038/s41467-021-21169-5},
archivePrefix = {arXiv},
       eprint = {2102.07602},
 primaryClass = {astro-ph.HE},
       adsurl = {https://ui.adsabs.harvard.edu/abs/2021NatCo..12.1025Y}
}

@ARTICLE{2025arXiv_He,
       author = {{He}, Han and {Long}, Yi and {You}, Bei and {Xie}, Fu-Guo and {Yan}, Zhen and {Zdziarski}, Andrzej A. and {Xu}, Sai-En},
        title = "{Daily fluctuations propagate damply through the accretion disk of Swift J1727.8-1613}",
      journal = {arXiv e-prints},
         year = 2025,
        month = aug,
          eid = {arXiv:2508.01384},
        pages = {arXiv:2508.01384},
          doi = {10.48550/arXiv.2508.01384},
archivePrefix = {arXiv},
       eprint = {2508.01384},
 primaryClass = {astro-ph.HE},
       adsurl = {https://ui.adsabs.harvard.edu/abs/2025arXiv250801384H}
}

@ARTICLE{2021ApJ_Zdziarski_hybrid,
       author = {{Zdziarski}, Andrzej A. and {Jourdain}, Elisabeth and {Lubi{\'n}ski}, Piotr and {Szanecki}, Micha{\l} and {Nied{\'z}wiecki}, Andrzej and {Veledina}, Alexandra and {Poutanen}, Juri and {Dzie{\l}ak}, Marta A. and {Roques}, Jean-Pierre},
        title = "{Hybrid Comptonization and Electron-Positron Pair Production in the Black-hole X-Ray Binary MAXI J1820+070}",
      journal = {\apjl},
         year = 2021,
        month = jun,
       volume = {914},
       number = {1},
          eid = {L5},
        pages = {L5},
          doi = {10.3847/2041-8213/ac0147},
archivePrefix = {arXiv},
       eprint = {2104.04316},
 primaryClass = {astro-ph.HE},
       adsurl = {https://ui.adsabs.harvard.edu/abs/2021ApJ...914L...5Z}
}

@ARTICLE{2017AJ_Madsen,
       author = {{Madsen}, Kristin K. and {Beardmore}, Andrew P. and {Forster}, Karl and {Guainazzi}, Matteo and {Marshall}, Herman L. and {Miller}, Eric D. and {Page}, Kim L. and {Stuhlinger}, Martin},
        title = "{IACHEC Cross-calibration of Chandra, NuSTAR, Swift, Suzaku, XMM-Newton with 3C 273 and PKS 2155-304}",
      journal = {\aj},
         year = 2017,
        month = jan,
       volume = {153},
       number = {1},
          eid = {2},
        pages = {2},
          doi = {10.3847/1538-3881/153/1/2},
archivePrefix = {arXiv},
       eprint = {1609.09032},
 primaryClass = {astro-ph.IM},
       adsurl = {https://ui.adsabs.harvard.edu/abs/2017AJ....153....2M}
}

@ARTICLE{2024A&A_Bouchet,
       author = {{Bouchet}, T. and {Rodriguez}, J. and {Cangemi}, F. and {Thalhammer}, P. and {Laurent}, P. and {Grinberg}, V. and {Wilms}, J. and {Pottschmidt}, K.},
        title = "{INTEGRAL/IBIS polarization detection in the hard and soft intermediate states of Swift J1727.8‒1613}",
      journal = {\aap},
         year = 2024,
        month = aug,
       volume = {688},
          eid = {L5},
        pages = {L5},
          doi = {10.1051/0004-6361/202450826},
archivePrefix = {arXiv},
       eprint = {2407.05871},
 primaryClass = {astro-ph.HE},
       adsurl = {https://ui.adsabs.harvard.edu/abs/2024A&A...688L...5B}
}

@INPROCEEDINGS{Belloni2016,
       author = {{Belloni}, Tomaso M. and {Motta}, Sara E.},
        title = "{Transient Black Hole Binaries}",
    booktitle = {Astrophysics of Black Holes: From Fundamental Aspects to Latest Developments},
         year = 2016,
       editor = {{Bambi}, Cosimo},
       series = {Astrophysics and Space Science Library},
       volume = {440},
        month = jan,
        pages = {61},
          doi = {10.1007/978-3-662-52859-4_2},
archivePrefix = {arXiv},
       eprint = {1603.07872},
 primaryClass = {astro-ph.HE},
       adsurl = {https://ui.adsabs.harvard.edu/abs/2016ASSL..440...61B}
}

@ARTICLE{2023ATelCastro-Tirado,
       author = {{Castro-Tirado}, A.~J. and {Sanchez-Ramirez}, R. and {Caballero-Garcia}, M.~D. and {Perez-Garcia}, I. and {Fernandez-Garcia}, E. and {Guziy}, S. and {Hu}, Y.-D. and {Blazek}, M. and {Hermelo}, I. and {Pinter}, V. and {Meintjes}, P.~J. and {van Heerden}, H.~J. and {Martin-Carrillo}, A. and {Hanlon}, L. and {Hiriart}, D. and {Lee}, W.~H. and {Carrasco-Garcia}, I.~M. and {Park}, I.~H. and {Gritsevich}, M. and {Castellon}, A. and {Perez del Pulgar}, C.~J. and {Reina}, A.},
        title = "{Optical spectroscopy of Swift J1727.8-1613 confirms a new low-mass X-ray binary hosting a black hole candidate}",
      journal = {The Astronomer's Telegram},
         year = 2023,
        month = aug,
       volume = {16208},
        pages = {1},
       adsurl = {https://ui.adsabs.harvard.edu/abs/2023ATel16208....1C}
}

@ARTICLE{2023ATelMiller-Jones,
       author = {{Miller-Jones}, J.~C.~A. and {Sivakoff}, G.~R. and {Bahramian}, A. and {Russell}, T.~D.},
        title = "{VLA radio detection of the new black hole X-ray binary candidate Swift J1727.8-1613}",
      journal = {The Astronomer's Telegram},
         year = 2023,
        month = aug,
       volume = {16211},
        pages = {1},
       adsurl = {https://ui.adsabs.harvard.edu/abs/2023ATel16211....1M}
}

@ARTICLE{2023ATelOConnor,
       author = {{O'Connor}, Brendan and {Hare}, Jeremy and {Younes}, George and {Gendreau}, Keith and {Arzoumanian}, Zaven and {Ferrara}, Elizabeth},
        title = "{NICER detection of Swift J1727.8-1613 (GRB 230824A)}",
      journal = {The Astronomer's Telegram},
         year = 2023,
        month = aug,
       volume = {16207},
        pages = {1},
       adsurl = {https://ui.adsabs.harvard.edu/abs/2023ATel16207....1O}
}

@ARTICLE{2025A&AMataSanchez,
       author = {{Mata S{\'a}nchez}, D. and {Torres}, M.~A.~P. and {Casares}, J. and {Mu{\~n}oz-Darias}, T. and {Armas Padilla}, M. and {Yanes-Rizo}, I.~V.},
        title = "{Dynamical confirmation of a black hole in the X-ray transient Swift J1727.8{\ensuremath{-}}1613}",
      journal = {\aap},
         year = 2025,
        month = jan,
       volume = {693},
          eid = {A129},
        pages = {A129},
          doi = {10.1051/0004-6361/202451960},
archivePrefix = {arXiv},
       eprint = {2408.13310},
 primaryClass = {astro-ph.HE},
       adsurl = {https://ui.adsabs.harvard.edu/abs/2025A&A...693A.129M}
}

@ARTICLE{Fender2009,
       author = {{Fender}, R.~P. and {Homan}, J. and {Belloni}, T.~M.},
        title = "{Jets from black hole X-ray binaries: testing, refining and extending empirical models for the coupling to X-rays}",
      journal = {\mnras},
         year = 2009,
        month = jul,
       volume = {396},
       number = {3},
        pages = {1370-1382},
          doi = {10.1111/j.1365-2966.2009.14841.x},
archivePrefix = {arXiv},
       eprint = {0903.5166},
 primaryClass = {astro-ph.HE},
       adsurl = {https://ui.adsabs.harvard.edu/abs/2009MNRAS.396.1370F}
}

@ARTICLE{Droulans2010,
       author = {{Droulans}, R. and {Belmont}, R. and {Malzac}, J. and {Jourdain}, E.},
        title = "{Variability and Spectral Modeling of the Hard X-ray Emission of GX 339-4 in a Bright Low/Hard State}",
      journal = {\apj},
         year = 2010,
        month = jul,
       volume = {717},
       number = {2},
        pages = {1022-1036},
          doi = {10.1088/0004-637X/717/2/1022},
archivePrefix = {arXiv},
       eprint = {1005.3195},
 primaryClass = {astro-ph.HE},
       adsurl = {https://ui.adsabs.harvard.edu/abs/2010ApJ...717.1022D}
}

@ARTICLE{McConnell2002ApJ,
       author = {{McConnell}, M.~L. and {Zdziarski}, A.~A. and {Bennett}, K. and {Bloemen}, H. and {Collmar}, W. and {Hermsen}, W. and {Kuiper}, L. and {Paciesas}, W. and {Phlips}, B.~F. and {Poutanen}, J. and {Ryan}, J.~M. and {Sch{\"o}nfelder}, V. and {Steinle}, H. and {Strong}, A.~W.},
        title = "{The Soft Gamma-Ray Spectral Variability of Cygnus X-1}",
      journal = {\apj},
         year = 2002,
        month = jun,
       volume = {572},
       number = {2},
        pages = {984-995},
          doi = {10.1086/340436},
archivePrefix = {arXiv},
       eprint = {astro-ph/0112326},
 primaryClass = {astro-ph},
       adsurl = {https://ui.adsabs.harvard.edu/abs/2002ApJ...572..984M}
}

@ARTICLE{Wood2025,
       author = {{Wood}, Callan M. and {Miller-Jones}, James C.~A. and {Bahramian}, Arash and {Tingay}, Steven J. and {Liu}, He-Xin and {Altamirano}, Diego and {Fender}, Rob and {K{\"o}rding}, Elmar and {Maitra}, Dipankar and {Markoff}, Sera and {Russell}, David M. and {Russell}, Thomas D. and {Sarazin}, Craig L. and {Sivakoff}, Gregory R. and {Soria}, Roberto and {Tetarenko}, Alexandra J. and {Tudose}, Valeriu},
        title = "{The Ejection of Transient Jets in Swift J1727.8{\ensuremath{-}}1613 Revealed by Time-dependent Visibility Modeling}",
      journal = {\apjl},
         year = 2025,
        month = may,
       volume = {984},
       number = {2},
          eid = {L53},
        pages = {L53},
          doi = {10.3847/2041-8213/adc9b3},
archivePrefix = {arXiv},
       eprint = {2503.03073},
 primaryClass = {astro-ph.HE},
       adsurl = {https://ui.adsabs.harvard.edu/abs/2025ApJ...984L..53W}
}

@ARTICLE{Yang2024,
       author = {{Yang}, Zi-Xu and {Zhang}, Liang and {Zhang}, Shuang-Nan and {Tao}, Lian and {Zhang}, Shu and {Ma}, Ruican and {Bu}, Qing-Cui and {Huang}, Yue and {Liu}, He-Xin and {Yu}, Wei and {Xiao}, Guangcheng and {Wang}, Peng-Ju and {Feng}, Hua and {Song}, Li-Ming and {Ma}, Xiang and {Ge}, Mingyu and {Zhao}, Qing-Chang and {Qu}, Jin-Lu},
        title = "{A Timing View of the Additional High-energy Spectral Component Discovered in the Black Hole Candidate Swift J1727.8-1613}",
      journal = {\apjl},
         year = 2024,
        month = aug,
       volume = {970},
       number = {2},
          eid = {L33},
        pages = {L33},
          doi = {10.3847/2041-8213/ad60bd},
archivePrefix = {arXiv},
       eprint = {2407.05236},
 primaryClass = {astro-ph.HE},
       adsurl = {https://ui.adsabs.harvard.edu/abs/2024ApJ...970L..33Y}
}

@ARTICLE{Hughes2025,
       author = {{Hughes}, Andrew K. and {Carotenuto}, Francesco and {Russell}, Thomas D. and {Tetarenko}, Alexandra J. and {Miller-Jones}, James C.~A. and {Bahramian}, Arash and {Bright}, Joe S. and {Cowie}, Fraser J. and {Fender}, Rob and {Gurwell}, Mark A. and {Khaulsay}, Jasvinderjit K. and {Kirby}, Anastasia and {Jones}, Serena and {Lescure}, Elodie and {McCollough}, Michael and {Plotkin}, Richard M. and {Rao}, Ramprasad and {Vrtilek}, Saeqa D. and {Williams-Baldwin}, David R.~A. and {Wood}, Callan M. and {Sivakoff}, Gregory R. and {Altamirano}, Diego and {Casella}, Piergiorgio and {Corbel}, St{\'e}phane and {DeBoer}, David R. and {Del Santo}, Melania and {Echibur{\'u}-Trujillo}, Constanza and {Farah}, Wael and {Gandhi}, Poshak and {Koljonen}, Karri I.~I. and {Maccarone}, Thomas and {Matthews}, James H. and {Markoff}, Sera B. and {Pollak}, Alexander W. and {Russell}, David M. and {Saikia}, Payaswini and {Castro Segura}, Noel and {Shaw}, Aarran W. and {Siemion}, Andrew and {Soria}, Roberto and {Tomsick}, John A. and {van den Eijnden}, Jakob},
        title = "{Comprehensive Radio Monitoring of the Black Hole X-Ray Binary Swift J1727.8{\ensuremath{-}}1613 during Its 2023─2024 Outburst}",
      journal = {\apj},
         year = 2025,
        month = jul,
       volume = {988},
       number = {1},
          eid = {109},
        pages = {109},
          doi = {10.3847/1538-4357/ade2e6},
archivePrefix = {arXiv},
       eprint = {2506.07798},
 primaryClass = {astro-ph.HE},
       adsurl = {https://ui.adsabs.harvard.edu/abs/2025ApJ...988..109H}
}

@ARTICLE{Zdziarski2017,
       author = {{Zdziarski}, Andrzej A. and {Malyshev}, Denys and {Chernyakova}, Maria and {Pooley}, Guy G.},
        title = "{High-energy gamma-rays from Cyg X-1}",
      journal = {\mnras},
         year = 2017,
        month = nov,
       volume = {471},
       number = {3},
        pages = {3657-3667},
          doi = {10.1093/mnras/stx1846},
archivePrefix = {arXiv},
       eprint = {1607.05059},
 primaryClass = {astro-ph.HE},
       adsurl = {https://ui.adsabs.harvard.edu/abs/2017MNRAS.471.3657Z}
}

@ARTICLE{Liao2025,
       author = {{Liao}, Jie and {Chang}, Ning and {Cui}, Lang and {Jiang}, Pengfei and {Mou}, Didong and {Huang}, Yongfeng and {An}, Tao and {Ho}, Luis C. and {Feng}, Hua and {Fu}, Yu-Cong and {Cao}, Hongmin and {Tripathi}, Ashutosh and {Liu}, Xiang},
        title = "{Tracking the Jet-like Corona of Black Hole Swift J1727.8{\ensuremath{-}}1613 during a Flare State through Type-C Quasiperiodic Oscillations}",
      journal = {\apj},
         year = 2025,
        month = jun,
       volume = {986},
       number = {1},
          eid = {3},
        pages = {3},
          doi = {10.3847/1538-4357/add264},
archivePrefix = {arXiv},
       eprint = {2410.06574},
 primaryClass = {astro-ph.HE},
       adsurl = {https://ui.adsabs.harvard.edu/abs/2025ApJ...986....3L}
}

@ARTICLE{Zdziarski2025,
       author = {{Zdziarski}, Andrzej A. and {Wood}, Callan M. and {Carotenuto}, Francesco},
        title = "{A Novel Method of Modeling Extended Emission of Compact Jets: Application to Swift J1727.8{\ensuremath{-}}1613}",
      journal = {\apjl},
         year = 2025,
        month = jun,
       volume = {986},
       number = {2},
          eid = {L35},
        pages = {L35},
          doi = {10.3847/2041-8213/ade13b},
archivePrefix = {arXiv},
       eprint = {2504.20962},
 primaryClass = {astro-ph.HE},
       adsurl = {https://ui.adsabs.harvard.edu/abs/2025ApJ...986L..35Z}
}

@ARTICLE{Shui2024,
       author = {{Shui}, Qing-Cang and {Zhang}, Shu and {Peng}, Jing-Qiang and {Zhang}, Shuang-Nan and {Chen}, Yu-Peng and {Ji}, Long and {Kong}, Ling-Da and {Feng}, Hua and {Yu}, Zhuo-Li and {Wang}, Peng-Ju and {Chang}, Zhi and {Yin}, Hong-Xing and {Qu}, Jin-Lu and {Tao}, Lian and {Ge}, Ming-Yu and {Zhang}, Liang and {Li}, Jian},
        title = "{Phase-resolved Spectroscopy of Low-frequency Quasiperiodic Oscillations from the Newly Discovered Black Hole X-Ray Binary Swift J1727.8-1613}",
      journal = {\apj},
         year = 2024,
        month = sep,
       volume = {973},
       number = {1},
          eid = {59},
        pages = {59},
          doi = {10.3847/1538-4357/ad676a},
archivePrefix = {arXiv},
       eprint = {2407.18106},
 primaryClass = {astro-ph.HE},
       adsurl = {https://ui.adsabs.harvard.edu/abs/2024ApJ...973...59S}
}

@ARTICLE{Peng2025,
       author = {{Peng}, Jing-Qiang and {Zhang}, Shu and {Shui}, Qing-Cang and {Zhang}, Shuang-Nan and {Chen}, Yu-Peng},
        title = "{A possible jet and corona configuration for Swift J1727.8─1613 during the hard state}",
      journal = {Journal of High Energy Astrophysics},
         year = 2025,
        month = mar,
       volume = {45},
        pages = {316-324},
          doi = {10.1016/j.jheap.2025.01.003},
archivePrefix = {arXiv},
       eprint = {2503.04044},
 primaryClass = {astro-ph.HE},
       adsurl = {https://ui.adsabs.harvard.edu/abs/2025JHEAp..45..316P}
}

@ARTICLE{Caojia2025,
       author = {{Cao}, Jia-Ying and {Liao}, Jin-Yuan and {Zhang}, Shuang-Nan and {Feng}, Hua and {Qu}, Jin-Lu and {Zhang}, Liang and {Liu}, He-Xin and {Yu}, Wei and {Zhao}, Qing-Chang and {Peng}, Jing-Qiang and {Ge}, Ming-Yu and {Tao}, Lian and {Xu}, Yan-Jun and {Zhang}, Shu and {Yang}, Zi-Xu},
        title = "{Spectral Analysis of the X-Ray Flares in the 2023 Outburst of the New Black Binary Transient Swift J1727.8─1613 Observed with Insight-HXMT}",
      journal = {\apj},
         year = 2025,
        month = apr,
       volume = {983},
       number = {1},
          eid = {23},
        pages = {23},
          doi = {10.3847/1538-4357/adbd0f},
archivePrefix = {arXiv},
       eprint = {2503.05411},
 primaryClass = {astro-ph.HE},
       adsurl = {https://ui.adsabs.harvard.edu/abs/2025ApJ...983...23C}
}

@ARTICLE{Fender2004,
       author = {{Fender}, R.~P. and {Belloni}, T.~M. and {Gallo}, E.},
        title = "{Towards a unified model for black hole X-ray binary jets}",
      journal = {\mnras},
         year = 2004,
        month = dec,
       volume = {355},
       number = {4},
        pages = {1105-1118},
          doi = {10.1111/j.1365-2966.2004.08384.x},
archivePrefix = {arXiv},
       eprint = {astro-ph/0409360},
 primaryClass = {astro-ph},
       adsurl = {https://ui.adsabs.harvard.edu/abs/2004MNRAS.355.1105F}
}

@ARTICLE{Fender2001,
       author = {{Fender}, R.~P.},
        title = "{Powerful jets from black hole X-ray binaries in low/hard X-ray states}",
      journal = {\mnras},
         year = 2001,
        month = mar,
       volume = {322},
       number = {1},
        pages = {31-42},
          doi = {10.1046/j.1365-8711.2001.04080.x},
archivePrefix = {arXiv},
       eprint = {astro-ph/0008447},
 primaryClass = {astro-ph},
       adsurl = {https://ui.adsabs.harvard.edu/abs/2001MNRAS.322...31F}
}
\bibliographystyle{aasjournal}

\end{document}